\documentclass[10pt,a4wide]{article}
\usepackage{newlfont}
\usepackage{amsmath}
\usepackage{graphics}
\usepackage{float}
\usepackage{graphicx}
\usepackage{epsfig}

\begin{document}
\title{Dynamical implications of gluonic excitations in \\meson-meson systems}
\author{Nosheen Akbar \thanks{e mail: noshinakbar@yahoo.com}\quad,
Bilal Masud \thanks{e mail: bilalmasud@chep.pu.edu.pk} \\
\textit{Centre For High Energy Physics, University of the Punjab University, Lahore-54590, Pakistan.}}
\date{}
\maketitle

\begin{abstract}
We study meson-meson interactions using an extended
$q^2\bar{q}^2(g)$ basis that allows calculating
coupling of an ordinary meson-meson system to a hybrid-hybrid one.
We use a potential model matrix in this extended basis which at
quark level is known to provide a good fit to numerical simulations
of a $q^2\bar{q}^2$ system in pure gluonic theory for static quarks
in a selection of geometries. We use a combination of resonating
group method formalism and Born approximation to include the quark
motion using wave functions of a $q\bar{q}$ potential within a cluster. This potential is taken to
be quadratic for ground states and has an additional smeared
$\frac{1}{r}$ (Gaussian) for the matrix elements between hybrid mesons. For the
parameters of this potential, we use values chosen to 1) minimize the error resulting from our use of a quadratic potential and 2) best fit the lattice data for differences
of $\Sigma_{g}$ and $\Pi_{u}$ configurations of the gluonic field
between a quark and an antiquark. At the quark (static) level, including the
gluonic excitations was noted to partially replace the need for
introducing many-body terms in a multi-quark potential. We study
how successful such a replacement is at the (dynamical) hadronic level of
relevance to actual hard experiments. Thus we study effects of both gluonic
excitations and many-body terms on mesonic transition amplitudes and the energy shifts resulting from the second order perturbation theory (i.e. from the respective hadron loops). The study suggests introducing both energy and orbital excitations in wave functions of scalar mesons that are modelled as meson-meson molecules or are supposed to have a
meson-meson component in their wave functions.
\end{abstract}

\maketitle
\begin{center}
\section*{1. INTRODUCTION}
\end{center}

Given the availability of both the numbers generated by lattice
simulations of quantum chromodynamics and continuum models of the
hadronic systems, an effective use of the numbers could be to
improve the models through constraints of getting a least chisquare
difference with the numbers for the corresponding discrete quarks
and antiquarks configurations. Such lattice-improved models can then
be reasonably used for all spatial configurations to eventually give
dynamical predictions for experimentally measurable quantities like
meson masses, meson-meson bindings and cross-sections and shifts
(polarization potentials) to meson masses arising through
meson-meson loops, etc. For one pair of quark and antiquark, a well
established such use of lattice results is substituting in a
Schr$\ddot{\textrm{o}}$dinger equation a Coulombic-plus-linear
quark-antiquark potential supported by lattice QCD calculations (see
ref. \cite{Bali1,Bali2,Alexandrou} and others) for the ground state of the
gluonic field between a quark and antiquark. Now that lattice
results for excited state of the gluonic field are also available
for years, even some dynamical uses of excited state gluonic field
potentials have been worked out~\cite{Morning1,Morning2,Nosheen}.

Uses worked out by others
are either limited to numerical calculations without an explicitly
written excited state gluonic field potential or the potential used
originates from flux tube~\cite{Isgur85} or string
models~\cite{string1,string2}. Each of these approaches has its usefulness.
What we add to this series of works is ourselves writing an analytical quadratic plus exponentially
falling expression for the excited state gluonic field potential
between a quark and antiquark and fit its parameters to the lattice
data for the excited state gluonic field values available for
discrete quark antiquark separations in~\cite{Morningstar03}.
This is reported in our previous work~\cite{Nosheen} as well. This work of us actually suggests and evaluates few other expressions for the excited state gluonic field potential as well. But the dynamical applications in it are present for a system whose valence
quark contents are limited to one quark and one antiquark.

Through the present paper, we extend work on the dynamical implications of gluonic excitations to multiquarks that can be composed to more than one hadronic subclusters. For this extension, we need quark-level potentials that can model the more complicated gluonic field of this multiquark system. For this we combine the modelling of the spatial distribution of this gluonic field reported in ref.~\cite{P. Pennanen98} (along with its fits to the continuum limits of the corresponding lattice simulations) with a realistic three colour structure. We have to do this combination because using all three colours the direct lattice simulations of {\em all} the Wilson loops relevant to a two quark two antiquark system are perhaps limited to ref.~\cite{V2005}
and those mentioned in refs.[9-11] of ref.~\cite{Green99}.
These works use a basis for a $q^2\bar{q}^2$ system that is truncated to the ground state of the glounic field. In comparison ref.~\cite{P. Pennanen98} extends the basis to include the gluonic excitations and its description of the spatial distribution of the $q^2\bar{q}^2$ glounic field is more complete. But the lattice simulations in ref.~\cite{P. Pennanen98} were carried out in a two-colour approximation to save computer time.

Our purpose here is to take advantage of the relatively
complete basis and spatial distribution models of ref.~\cite{P. Pennanen98} but using all three colours in the quark-level potential we use for our calculations of the dynamical hadron-level implications for a meson-meson system. In changing a two colour based model to a full three colour one,
the number of colour basis states that interact (for any the spatial configuration) remain the same (see eqs. A.1 and A.3 of ref.~\cite{bridges},
along with eqs. 5.1 to 5.4 and fig. 5.1 of ref.~\cite{thesis}) and
we had to essentially only replace some colour overlap factors with proper $SU(3)_c$ values as elements of matrices of the same order. But the inter-quark elementary potential of ref.~\cite{P. Pennanen98} turned out to be problematic for including the quark motion for the hadron level implications and thus we had to replace their numerically fitted
$0.562+0.0696r-\frac{0.255}{r}-\frac{0.045}{r^2}$ ground state quark
antiquark potential by a constant
plus quadratic confining potential term and the {\em additional}
$\frac{\pi}{r}-\frac{4.24}{r^2}+\frac{3.983}{r^4}$ potential for the
 gluonic excitation by one of the form
$A\textrm{\text{exp}}(-B r^2)$.
As written in ref.~\cite{Nosheen}, only $A\textrm{\text{exp}}(-B r^2)$ can be used in solving the integrals of our present work analytically although few other forms for the gluonic potential are also suggested in ref.~\cite{Nosheen} with less $\chi^2$ as compared to $A \textrm{exp}{- B r^2}$. Only then we are able to perform a full meson level dynamical calculations for transition
amplitudes from one set of quark-antiquark clusterings (mesons) to
the other. Using these amplitudes we are also able to study
certain properties of the polarization potentials for a meson-meson system.

We had to use simpler interquark potentials
that can be symbolically integrated at a later stage, after necessary multiplications by wave functions of $qq\bar{q}\bar{q}$ positions, to complete the adiabatic-approximation-based treatment of a $qq\bar{q}\bar{q}$ system mentioned below.
Being not limited by such demands of later integrations, the form and then parameters values of the
continuum $qq\bar{q}\bar{q}$ model proposed in ref.~\cite{P. Pennanen98}
were chosen to simply minimize
\begin{equation}
\chi^{2}_{A} = \frac{1}{N (G)}\sum^{N(G)}_{i=1} (E_{i}-M_{i})^{2}/\triangle E_{i},\label{i}
\end{equation}
where $N (G)$ is the number of data points for geometry G.
The $qq\bar{q}\bar{q}$ geometries in ref.~\cite{P. Pennanen98} numerically
worked on were (quarks at the corners of) squares, rectangles,
tetrahedra and some other less symmetric geometries $Q,N,P$ and $L$
(linear). For each data point $i$ , the lattice energy $E_{i}$ was
extracted by solving the following eigenvalue equation (see eq. 2 of ref.~\cite{Green96})
\begin{equation}
W^{T}_{ikl} a^{T}_{il} = \lambda^{(T)}_i W^{T-1}_{ikl}a^{T}_{il}\label{ii}
\end{equation}
for $\lambda^{(T)}_i$ that approaches to ${\text {\text{exp}}}
(-V_i)$ as (Euclidean time) $T \rightarrow \infty$, and then subtracting the energy of two
separated $q\bar{q}$ clusters from the $V_{i}$ to get the
lattice-generated $qq\bar{q}\bar{q}$ binding energy $E_{i}$ for the
data point.  The values of the $\lambda^{(T)}_i$, and thus of
$E_i$, depend only on the numerical values of the elements
$W^{T}_{ikl}$ of the matrix of the Wilson loops. The
values of $k$ and $l$ depend on the number of Wilson loops
evaluated; for the $qq\bar{q}\bar{q}$ system these were taken to be
1 and 2. Two of the corresponding Wilson operators (whose
vacuum expectation values are the Wilson loops $W^{T}_{kl}$) are
shown for example in Fig. 1.5 of ~\cite{bridges}.
Knowing the Wilson loops, the procedure of getting $V_{i}$ can be
found for example in eqs. 4, 11, 12 and 15 of ref.~\cite{Green96}. The arguments
for continuum limits being achieved before extracting $E_i$ are given in ref.~\cite{pennanen}.

The  $M_{i}$ in eq.\eqref{i}
are obtained by subtracting the energy of two separated $q\bar{q}$
clusters from the eigenvalues of a matrix obtained through a {\em
model} of the $qq\bar{q}\bar{q}$ system. For this, the model has to
give a basis and an operator $\widehat{V}$
whose representation with respect to the basis gives a potential matrix $V$.
$M_i$ are obtained by setting the determinant of $V-(M_i +
V_{11})N$ equal to zero, with $V_{11}$ being the energy of two
separated $q\bar{q}$ clusters and $N$ the (overlap) matrix of
an identity operator in the basis. Searching for the model, the
simplest way to extend a two-particle potential model to a few-body
is to use the potential for each pair of particles in the few-body
system and simply add up such two-body potentials. This approach has
been successful in atomic and many-nucleon systems; the
corresponding two-body interaction being described by Coulombic and
Yukawa potential, for example. For a hadron (or a system of hadrons)
composed of many quarks, antiquarks and the gluonic field, the
lowest order perturbative Feynman amplitudes are of this sum-of-pair-wise form. Though  Feynman diagrams themselves
become impractical for typical hadronic energies because of larger
couplings, models have been tried which simply replace the two-body
Coulombic potential (essentially a Fourier transform of the Gluonic propagator) by more
general Coulombic-plus-linear form; see ref. \cite{T. Barnes92}.
This approach is not free of problems; for example  it leads to
inverse power van der Waals' potentials~\cite{Greenberg81} between
separated colour-singlet hadrons which are in contradiction with
experimental data. But this model has many phenomenological
successes and it is worthwhile inquiring if

 1) it provides a basis and operator to generate a potential matrix,  and

 2) how good is the chisquare if the eigenvalues of the resulting matrix are used as $M_i$ in
eq.\eqref{i}.

\noindent The answer provided by ref.~\cite{P. Pennanen98}
and earlier related works is that the model does generate a matrix
of the required kind. But the resulting chisquare, defined by
eq.\eqref{i}, is too bad; see Fig. 4 of ref.\cite{Green1}. To refine
the model we can improve the basis beyond the $\{|1\rangle,
|2\rangle, |3\rangle\}$ defined as ~\cite{P. Pennanen98,P.
PennanenB}
\begin{equation}
|1\rangle = (q_{1}q_{3})(q_{2}q_{4}),\quad |2\rangle = (q_{1}q_{4})(q_{2}q_{3}), \textrm{ and} \quad |3\rangle = (q_{1}q_{2})(q_{3}q_{4}),\label{i3}
\end{equation}
and the operator beyond
\begin{equation}
H = - \sum^{^{4}}_{i = 1}\big[m_{i} + \frac{\widehat{P}_{i}^{2}}{2 m_{i}}\big] + \sum_{i<j} v_{ij}\textbf{F}_{i}.\textbf{F}_{j},\label{i4}
\end{equation}
with $v_{ij}$ being is the potential energy of a $q\bar{q}$ pair
with the gluonic field between them in the ground state. Or both the
basis and the operator can be improved. What ref.~\cite{P.
Pennanen98} does is to improve directly the matrix (representation)
after writing down the underlying basis. The authors do this in a number of
ways. One model, termed model $\textrm{II}$, uses the same
$\{|1\rangle,|2\rangle,|3\rangle\}$ basis but multiplies the
off-diagonal elements of the overlap and potential energy matrices
(that is, the representations of the identity and potential
operators $$\sum_{i<j} v_{ij}\textbf{F}_{i}.\textbf{F}_{j}$$
respectively) by a few-body gluonic field overlap factor
$f=\text{exp}(-b_{s} k_{f} S )$ with $b_s$ as the tension of the
string connecting a quark with an antiquark, $S$ the area of a
surface bounded by external four lines connecting two quarks and two
antiquark and $k_f=0.5$ approximately; theoretical arguments suggest
$S$ should be the area of the corresponding minimal surface, though
in ref.~\cite{P. Pennanen98}  half of a sum of four triangles was
used for numerical convenience. At the quark level, this model $\textrm{II}$ was
noted to much reduce the chisquare of
eq.\eqref{i}. This model has been worked out in ~\cite{B. Masud91,B.
Masud94,Imran} till meson-level transition amplitudes. The dynamical
calculations require a kinetic energy term as well. As this is taken to be apart from some technical considerations of hermicity,
proportional to the overlap matrix and hence its off-diagonal
elements are also multiplied by the overlap $f$ factor. Thus $f$
provides one parametrization that connects QCD simulations with hard
experiments.

But model $\textrm{II}$ is not the best model of ref.~\cite{P.
Pennanen98}; the paper continues to then improve the basis by
including the gluonic excitations as well. That is, it extends the
$\{|1\rangle,|2\rangle,|3\rangle\}$  basis by including the states
\begin{align}
|1^{\star}\rangle = (q_{1}q_{3})_{g}(q_{2}q_{4})_{g},\quad |2^{\star}\rangle = (q_{1}q_{4})_{g} (q_{2}q_{3})_{g}, \textrm{ and} \quad |3^{\star}\rangle =
(q_{1}q_{2})_{g} (q_{3}q_{4})_{g}.\label{i5}
\end{align}
Here $(q_{1}q_{3})_{g}$ denotes a state where the gluon field is
{\em excited} to the lowest state. (The excited states of gluonic
field can, for example, be seen in the QCD numerical simulations;
see ref.\cite{Morningstar98,Morningstar03} and others). When the overlap,
potential and kinetic energy matrices are written in this extended
basis, their order increases to $6\times 6$ rather than previous $3
\times 3$. If in addition, we introduce many body terms in this
extended model, new kind of gluonic field overlap factors ($f^{a}$, $f^{c}$) appear in the off diagonal terms resulting in what ref.~\cite{P. Pennanen98} terms model
$\textrm{III}$ giving the least chisqure
in ref.~\cite{P. Pennanen98}; (see eq.\eqref{e6} below); our truncation to $4 \times 4$ matrices is explained
before this equation. As mentioned above, the purpose of our present paper is to work out
this improved model $\textrm{III}$ of ref.~\cite{P. Pennanen98} till
the meson-meson scattering amplitudes and energy shifts. As this
improved model $\textrm{III}$
includes the gluonic excitations, it consider transitions from three
 ground state quark states to the
 ones having gluonic excitations. And by adding to it the quark motion (wave functions) to reach the hadron level, we are now able to study transitions from ground state meson-meson systems to hybrid-hybrid systems.

A worth-mentioning aspect we have studied is the hadron-level
implications of the differences of the gluonic-excitation-including
model $\textrm{III}$  and the sum-of-pair-wise approach. The
quark-level work in ref.~\cite{P. Pennanen98} can be interpreted to
mean that with suitable extensions and modifications something like
a sum-of-pairs approach can be a good approximation for a tetraquark system as well. It states "At the shortest distances, upto
about 0.2fm, perturbation theory is reasonable with the binding being given mainly by the $|1\rangle$, $|2\rangle$, and $|3\rangle$ states inter-acting simply through the two-quark potentials with little effect from four-quark potentials". However, ref.~\cite{P.
Pennanen98} states, "for large inter-quark distance (greater than 0.5 fermi), quark-pair creation
can no longer be neglected. However, in the
intermediate energy range, from about 0.2 to 0.5 fm, the four-quark
potentials act in such a way as to reduce the effect of the
$|1\rangle$, $|2\rangle$, and $|3\rangle$ states so that the binding
is dominated by the $|1^{\star}\rangle$, $|2^{\star}\rangle$, and
$|3^{\star}\rangle$ states, which now interact among themselves
again simply through the two-quark potentials with little effect
from four-quark potentials." This suggests that models involving
only two-quark potentials could be justified provided excited gluon
states (such as $|1^{\star}\rangle$, $|2^{\star}\rangle$,
$|3^{\star}\rangle$) are included on the same footing as the
standard states $|1\rangle$, $|2\rangle$,$|3\rangle$. We have checked if such features survive at the experimentally
meaningful hadron level, by comparing the dynamical implications of

(1) a model extended to the gluonic excitations but otherwise
sharing many features with the sum-of-pair-wise approach, with

(2) a model that includes explicit many-body terms but does not include
gluonic excitations.

\noindent Thus we report if after including the gluonic
excitations a sum-of-two-body potential model can replace to some
extent many-body potential terms in a tetraquark system {\em even at
a hadronic level.} Specifically, we have calculated in both kind of models meson-meson transition amplitudes $T_{ij}$ from ($i$) a ground state meson-meson clustering to ($j$) a different ground-state clustering and to a clustering of gluonic-excited mesons. $T_{ij}$ are elements of the meson-meson scattering theory $T$-matrix, can be termed as phase shifts, transition
potentials or meson-meson coupling, and their absolute squares give meson-meson differential cross sections \cite{T.
Barnes92}. Moreover, using these transition amplitudes in the second order
perturbation theory, we study shifts (in both kind of models) in a ground state meson-meson energy due to coupling to a different ground-state clustering and to an excited state meson-meson system (i.e. to {\em hybrid loops}).
These energy shifts are what are also termed polarization potentials ~\cite{wong}.

To reach the hadronic level, we have included the quark motion through quark wave functions. To find the quark
position dependence in multiquark systems, a number of methods are used such as variational method~\cite{Isgur83,Isgur90}, Born-order diagrams~\cite{T. Barnes92}, and resonating group
method~\cite{Gross73}. Variational approach is used by Weinstein and
Isgur to optimize a meson-meson wave function
in a quadratic ~\cite{Isgur83} and later coulomb plus linear~\cite{Isgur90} potential and a
hyperfine term combined with a sum-of-two-body approach. Then they projected the
meson-meson state onto free meson wave functions to estimate a
relative two meson wave function which gives the equivalent meson-meson potential and obtained the meson-meson phase shifts.
The similar results can be obtained by using Born-order quark exchange
diagrams ~\cite{T. Barnes92} in a non-relativistic potential model
to describe low energy scattering of $q\overline{q}$ mesons.

In the present paper, we have used a formalism of the
resonating method as used in~\cite{B. Masud91}. In the resonating
group method, the dependence on the internal co-ordinates of the
system is specified before solving the problem to integrate out the
degrees of freedom corresponding to the internal coordinates of
clusters of the system. At a later stage, because of the complexity
of the calculations we use a Born approximation
to specify the dependence on the vectors
connecting the centers of masses of our mesonic clusters as well. Moreover, we have not included
in our basis an explicit diquark-diantiquark state. In the weak
coupling limit such a state is a linear combination of the
meson-meson states and thus cannot be included in a basis. Away from
the weak coupling this can be included. But inclusion of a third clustering states
 in the basis did not affect the lattice simulations reported in ref.~\cite{P.
Pennanen98}. Later works ~\cite{Green99,V2005} using $SU(3)_c$ were done with only two clusterings.
Thus we have expanded our two-quark two antiquark
wave function in a basis that is limited to four meson-meson states: ground and excited states for each of the two
possible $q\bar{q}q\bar{q}$ clusterings.

In this exploratory work, we have taken all the constituent quark
masses to be same as that of a charm quark and we have taken all the
spin overlaps to be 1 without calculating them. Actually we have not studied the effects of the quark degree of freedom, not its orbital angular momentum as well; we want to
study only  dynamical effects of including the gluonic excitations in our
basis for the same quark configurations. In section 2, the potential model in the extended basis in the pure
gluonic theory is introduced for the static quarks. Basically, in
this section we tell where does the model of ref.~\cite{P. Pennanen98}
 fits in our
full scheme that incorporates the quark motion through a resonating
group method formalism meaning pre-specifying quark-antiquark
wave functions within $q\bar{q}$ clusters. The coupled integral
equations for the remaining inter-cluster wave function
$\chi_{K}(\textbf{R}_{K})$ are written in section 3. In section 4,
these integral equations are solved to calculate the transition
amplitudes and energy shifts we are studying.
 The numerical results for meson-meson system transition amplitudes and energy
shifts in both the above mentioned kinds of models
with concluding remarks are given in section 5. The partial wave analysis results of
transition from ground state to excited state gluonic field are also reported in section 5.

\section*{\begin{center}2. $q^{2}\overline{q}^{2}$ POTENTIAL MODEL (IN THE EXTENDED BASIS)\end{center}}

\qquad Using the adiabatic approximation, the total state vector of a
system containing two quarks, two anti-quarks and the gluonic field
between them can be written as sum of product of quark position
dependence function
$\phi_{K}(\textbf{r}_{1},\textbf{r}_{2},\textbf{r}_{\overline{3}},\textbf{r}_{\overline{4}})$
and the gluonic field state $\mid k \rangle_{g}$. (The gluonic state
$\mid k \rangle_{g}$ is defined as a state approaching to colour
state $\mid k \rangle_{c}$ in the limit of quark anti-quark
separation approaching to zero. Here
$k=1,2,3,1^{\star},2^{\star},3^{\star}$.)
\smallskip
The function
$\phi_{K}(\textbf{r}_{1},\textbf{r}_{2},\textbf{r}_{\overline{3}},\textbf{r}_{\overline{4}})$
can be written as
\begin{align*}
\phi_{K}(\textbf{r}_{1},\textbf{r}_{2},\textbf{r}_{\overline{3}},\textbf{r}_{\overline{4}}) =
\phi_{K}(\textbf{R}_{c},\textbf{R}_{K},\textbf{y}_{K},\textbf{z}_{K}),
\end{align*}
with $K = 1, 2, 3$. $R_{c}$ is the overall center of mass co-ordinate of the whole system.
\begin{figure}
\begin{center}
\epsfig{file=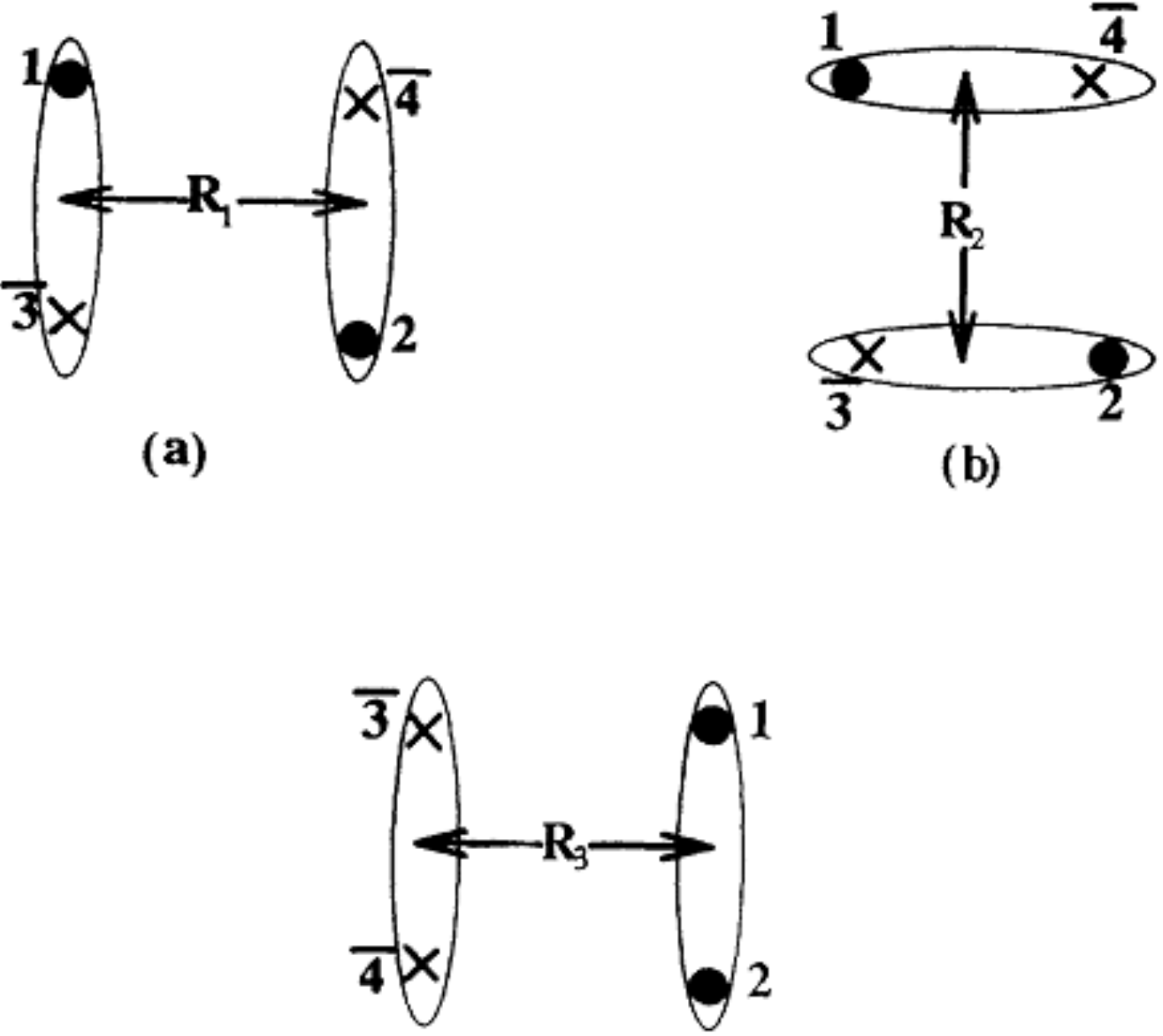,width=0.7\linewidth,clip=} \caption{(a) and (b) describes the meson-meson topologies and (c) describes diquark and diantiquark topology.}\label{1}
\end{center}
\end{figure}\\
With the notation of Fig. 1, the relative co-ordinates
$\textbf{R}_{1},\textbf{R}_{2},$ and $\textbf{R}_{3}$ are defined as
\begin{align*}
\textbf{R}_{1} = \frac{1}{2}(\textbf{r}_{1} + \textbf{r}_{\overline{3}} - \textbf{r}_{2} - \textbf{r}_{\overline{4}}),
\textbf{y}_{1} = \textbf{r}_{1} - \textbf{r}_{\overline{3}},\textbf{z}_{1} = \textbf{r}_{2} - \textbf{r}_{\overline{4}}\\
\textbf{R}_{2} = \frac{1}{2}(\textbf{r}_{1} + \textbf{r}_{\overline{4}} - \textbf{r}_{2} - \textbf{r}_{\overline{3}}),
\textbf{y}_{2} = \textbf{r}_{1} - \textbf{r}_{\overline{4}},\textbf{z}_{2} = \textbf{r}_{2} - \textbf{r}_{\overline{3}}\\
\textbf{R}_{3} = \frac{1}{2}(\textbf{r}_{1} + \textbf{r}_{2} - \textbf{r}_{\overline{3}} - \textbf{r}_{\overline{4}}), \textbf{y}_{3} =
\textbf{r}_{1} - \textbf{r}_{2},\textbf{z}_{3} = \textbf{r}_{\overline{3}} - \textbf{r}_{\overline{4}},
\end{align*}
$\textbf{R}_{1}$ being the vector joining the centers of mass of the
mesonic clusters $(1,\overline{3})$ and $(2,\overline{4})$;
similarly about $\textbf{R}_{2}$ and $\textbf{R}_{3}$. Now using the
resonating group method, the quark position dependence function can
be written as a product of function of known dependence on
$\textbf{R}_{c}$,$\textbf{y}_{K}$,$\textbf{z}_{K}$ and of unknown
dependence on $\textbf{R}_{K}$. i.e.
\begin{equation}
\phi_{K}(\textbf{R}_{c},\textbf{R}_{K},\textbf{y}_{K},\textbf{z}_{K}) = \psi_{c}(\textbf{R}_{c}) \chi_{K}(\textbf{R}_{K})
\psi_{K}(\textbf{y}_{K},\textbf{z}_{K}).\label{i6}
\end{equation}
Thus, the two quarks two antiquarks state vector can be written as
\begin{equation}
\begin{split}
\mid \Psi(q_{1}q_{2}q_{\overline{3}}q_{\overline{4}})
\rangle = \sum_{k}\mid
k\rangle_{g}\psi_{c}(\textbf{R}_{c})\chi_{k}(\textbf{R}_{K}) \psi_{k}(\textbf{y}_{K},\textbf{z}_{K}),\label{i7}
\end{split}
\end{equation}
where
\begin{align*}
\psi_{k}(\textbf{y}_{K},\textbf{z}_{K}) = \xi_{k}(\textbf{y}_{K}) \xi_{k}(\textbf{z}_{K}),
\end{align*}
$\xi_{k}(\textbf{y}_{K})$ and $\xi_{k}(\textbf{z}_{K})$ being the
normalized solutions of the Schr$\ddot{\textrm{o}}$dinger equation
for quadratic confining potential (written in eq.\eqref{e1}) for a pair of quark-anti-quark within a cluster.
We take, for the zero relative orbital momentum ($\ell$) of a quark w.r.t. the antiquark of the cluster,
\begin{equation}
\begin{split}
\xi_{K}(\textbf{y}_{K}) = \frac{1}{(2 \pi
d^{2})^{\frac{3}{4}}}\text{exp}( - \frac{y_{K}^{2}}{4 d^{2}}),\\
\xi_{K}(\textbf{z}_{K}) = \frac{1}{(2 \pi d^{2})^{\frac{3}{4}}} \text{exp}(- \frac{z_{K}^{2}}{4 d^{2}}).\label{i8}
\end{split}
\end{equation}
Here $d$ is the size of meson (detail is written after
eq.\eqref{e1}) and $m$ being the constitute quark mass. In our case,
$m$ is the mass of c-quark and equal to $ 1.4794$ GeV as used in ~\cite{T.
Barnes05}.

After writing the form of the wave vector, we describe our Hamiltonian,
starting with the limit when each gluonic field overlap factor $f =
f^{a} = f^{c} = 1$. In this limit, the Hamiltonian whose representation matrices in the
basis \{$|1\rangle ,|2\rangle ,|3\rangle ,|1^{\star}\rangle ,|2^{\star}\rangle ,|3^{\star}\rangle $\} would become those in ref.~\cite{P. Pennanen98} is
\begin{equation}
H = - \sum^{^{4}}_{i = 1}\big[m_{i} + \frac{\widehat{P}_{i}^{2}}{2 m_{i}}\big] + \sum_{i<j}( v_{ij} +
\epsilon \triangle v^{\star}_{ij})\textbf{F}_{i}.\textbf{F}_{j},\label{i9}
\end{equation}
where $v_{ij}$ is the potential energy of meson for the ground state
gluonic field and $\triangle v^{\star}_{ij}$ is the difference
between ground state and excited state gluonic field potential.
We take the kinetic energy in the non-relativistic limits. This limit is also used in a recent work by Vijande
ref.~\cite{Vijande} that deals with multiquark system (two quarks and two antiquarks) to study the spectrum using a string model for the potential. In ref.~\cite{P. Pennanen98}, potential energy matrix elements are
written so that the potential energy for each pair $ij$ is equal to
$v_{ij}$ for the matrix elements of the Hamiltonian between gluonic
ground states, and it is equal to $v_{ij} + \triangle
v^{\star}_{ij}$ for the matrix elements between the gluonic-excited
states. We have modeled these two forms by taking $\epsilon=0$ for
the ground state matrix elements and $\epsilon=1$ for the elements
between gluonic-excited states. For the elements between ground and
excited state gluonic field, the $\epsilon$ value that results from
their parameter $a_0$ being fitted to 4 (in their Table 1) is
surprisingly $2$ and not any value between 0 and 1. A possibility is
that this is a result of them taking the area $S$, we mentioned
above in our introduction, in the form of average of the sum of triangle areas instead of the
theoretically motivated minimal surface area. Thus we have somewhat
explored $\epsilon=\frac{1}{2}$ between 0 and 1 and $\epsilon=1$ in
addition to $\epsilon=2$ which we have mainly studied.

In the above equation, $\textbf{F}_{i}$ (operating on $i_{th}$
particle) has eight components $\textrm{F}^{\l}_{i}$ with ${\l}=1,2,3,.....,8$. Each component is equal to $
\frac{\lambda^{\l}_{i}}{2}$, where $\lambda^{\l}_{i}$ are the Gell-Mann
matrices. ${\l}$ is used in the superscript to avoid the possible confusion with index $i$. We used the potential with the colour structure of one gluon
exchange in the form given in ref.~\cite{P. Pennanen98}. With the
use of ground state potential $ v_{ij}$ in the realistic coulombic plus linear
form, it becomes impossible for us to solve the integral equations
appearing below in eqs.(\ref{e12}-\ref{e15}). Therefore we used the
parametrization of the static  pairwise two quark potential as
\begin{equation}
v_{ij} = C r^{2}_{ij} + \overline{C}, \qquad \textrm{with} \quad i, j=1, 2, \overline{3}, \overline{4}.\label{e1}
\end{equation}
In this simple harmonic potential, the parameters $C$ is related to size ($d$) of
wavefunction ($\xi_{k}$) through the relation $C = -\frac{3}{16}{\frac{1}{d^{4}}} = -\frac{3}{16}m \omega^{2}$ with $\omega = {\frac{1}{m d^{2}}}$, and for consistency of the diagonal term of the integral eq.(\ref{e12})
$\overline{C} = \frac{3}{8}(4 m + 3 \omega - 2 M)$ GeV~\cite{B. Masud91} with $M = 2
m = 2.9588$GeV being the mass of a charmonium meson. The parameter $d$
is chosen in such a way to
reduce the error resulting from a use of this quadratic potential
instead of the realistic one. The error may be both in the wave
functions of the distance between a quark and antiquark within
clusters and those of the vectors joining the centers of masses of the
two clusters. As for the first dependence, we found that the maximum of the
overlap integral of each of the wave functions $ \xi_{K}(\textbf{y}_{K})$ and $\xi_{K}(\textbf{z}_{K})$
of the quadratic potential and that of a more realistic coulombic plus linear potential
($-\frac{4}{3 r} \alpha_s + b_s r$) is $99 \%$ at $d = 1.16
\textrm{GeV}^{-1}$. (For the parameters of the realistic potential we used values $\alpha_{s} =
0.5461$~\cite{T. Barnes05} and $b_{s} = 0.1425$$\textrm{GeV}^2$~\cite{T.
Barnes05} for mesons composed of charm quarks.)
This overlap is shown in Fig. 2. A similar work was done in
ref.~\cite{T. Barnes96} for lighter quarks. They found that the
overlap wave function of SHO (quadratic potential) and that of
coulombic plus linear can be made as large as $99.4 \%$ with the
suitable adjustment of parameters.
\begin{figure}
\begin{center}
\epsfig{file=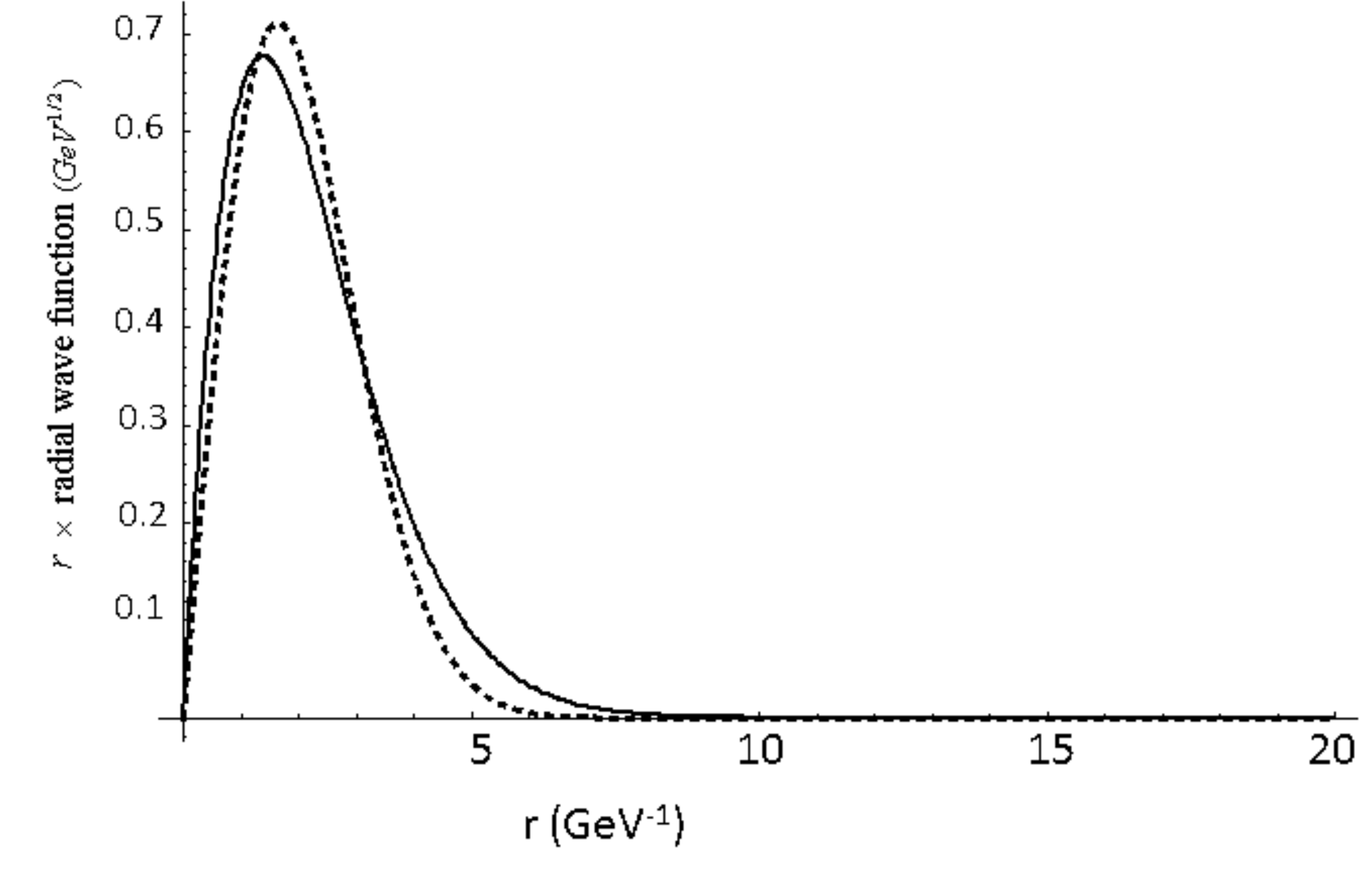,width=0.9\linewidth,clip=} \caption{Overlap of the (ground state) wave function of the realistic Coulombic plus linear potential (shown by solid line) and that of the quadratic potential (shown by dashed line).}\label{1}
\end{center}
\end{figure}

For the additional term in the $q\bar{q}$ potential for the gluonic
excitation, the usual flux tube ($\pi/r$~\cite{Isgur85}) or string
based analytical expressions become impractical for us, as mentioned
in the introduction. Thus for that we tried an $ans\ddot{a}tz$ of the form of
\begin{equation}
\triangle v^{\star}_{ij} = A e^{- B r_{ij}^{2}}.\label{e2}
\end{equation}
This gaussian gluonic potential ($A \text{exp}(- B r_{ij}^{2})$) is
a smeared form of $\frac{{\text constant}}{r}$ as written in appendix of ref.~\cite{Isgur90}. From the Fig. 3 of~\cite{Morningstar03}, we get the potential energy
difference $\varepsilon_{i}$ between ground and excited states for
different $\emph{r}$ values ($r_{i}$). We choose A and B for which
$\chi^{2}$ becomes minimum. $\chi^{2}$ is defined as
\begin{equation*}
\chi^{2} = \sum^{^{n}}_{_{i=1}}(\varepsilon_{i} - A \text{exp} [- B
r_{ij}^{2}])^{2},
\end{equation*}
with $n$  being the number of data points. This gives
\begin{equation*}
A = 1.8139 \textrm{GeV},\qquad B = 0.0657 \textrm{GeV}^{2}.
\end{equation*}
For finding the wave function corresponding to our total potential
$C r^{2}_{ij} + \overline{C} + A e^{- B r^{2}_{ij}}$, we used the
variational method with an $ans\ddot{a}tz$ wave function
\begin{equation}
\xi^{\star}_{K}(\textbf{y}_{K}) = n y_{K}^{2}\text{exp}(- p y_{K}^{2}).\label{e3}
\end{equation}
The normalization of this $\xi^{\star}_{K}(\textbf{y}_{K})$ w.r.t $y_{K}$ gives
\begin{align*}
n = (4 2^{\frac{3}{4}}p^{\frac{7}{4}})(15^{\frac{1}{2}}\pi^{\frac{3}{4}}).
\end{align*}
This leaves us with one variational parameter $p$ chosen to minimize
the expectation value of the two body Hamiltonian in the excited
state gluonic field wave function. This gave $p = 0.048 GeV^{2}$. For this value of $p$, the overlap of wave function of the quadratic potential
plus $\triangle v^{\star}_{ij}$ and that of coulombic plus linear
plus $\triangle v^{\star}_{ij}$ within a hybrid
cluster became $99.9 \%$. Both wave functions are shown in in Fig. 3.
Having much reduced the errors in the in-cluster factors of the
total wave function, the question remains how much the inter-cluster
factors of the (terms of the) total state vector are affected by our
use of convenient but not realistic $q\bar{q}$ potentials. For the
inter-cluster wave functions, eventually we use below in
eq.~\eqref{e16}  plane wave forms  which get their justifications from the validity of Born
approximation for our problem regardless of potential expressions we
use. This plane wave form has only one usual parameter (the wave
number) and eq.(C10) below relates its value for the ground as well as
excited state inter-cluster waves to the very good values of $d$ and
$p$ that almost give realistic ground state and excited state wave
functions within $q\bar{q}$ clusters. But the relations between the
inter-cluster wave numbers and the $d$ and $p$
assume a quadratic confinement and this may affect our numerical results but hopefully not at least
the qualitative features we are pointing out. Perhaps it is worth
mentioning here that properties of $q^{2}\overline{q}^{2}$ systems
were calculated using quadratic confinement in ref.~\cite{Isgur83},
and then with the realistic potential in ref.~\cite{Isgur90} and
both the works favoured the existence of meson-meson molecules.
\begin{figure}
\begin{center}
\epsfig{file=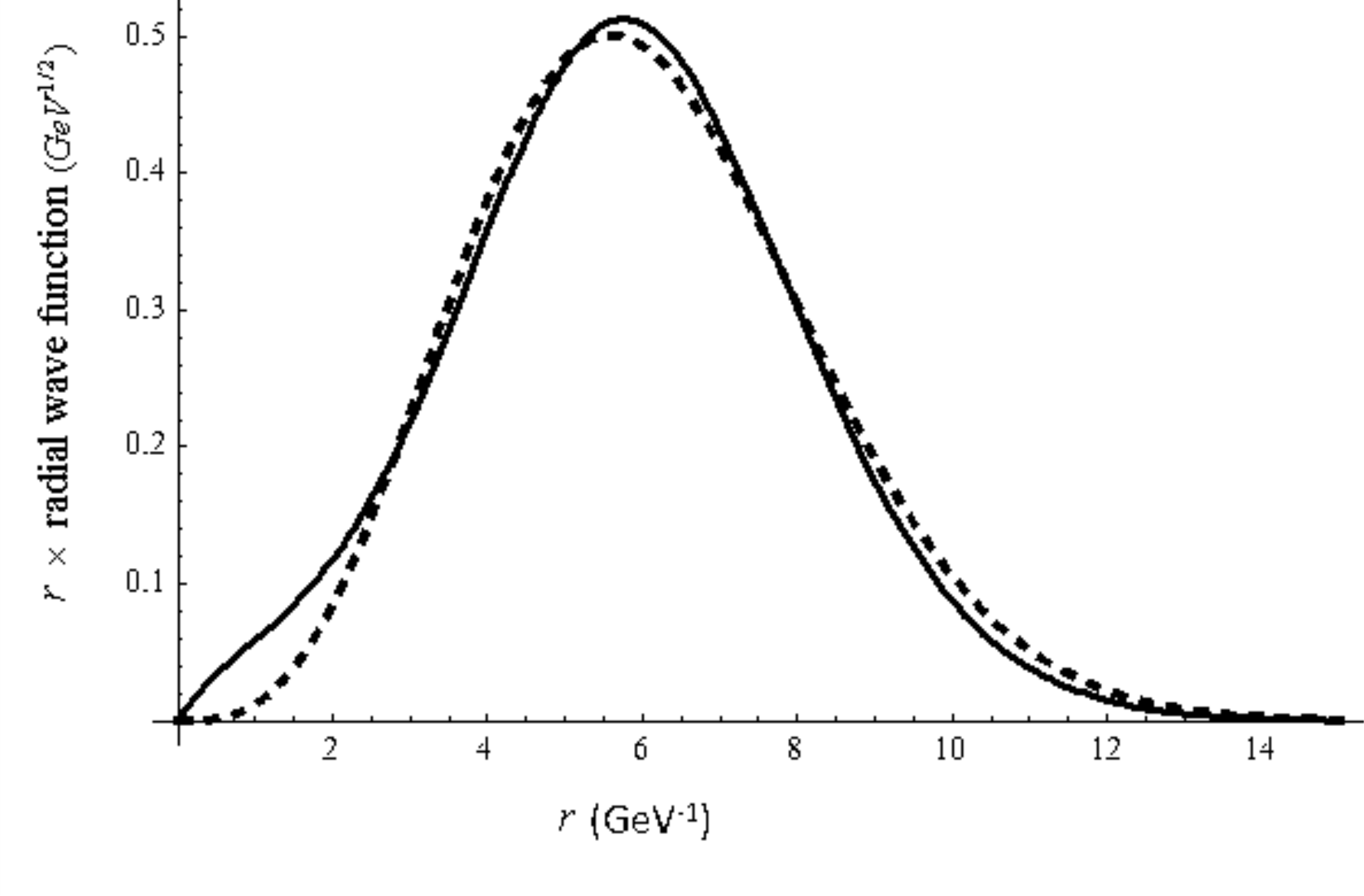,width=0.9\linewidth,clip=} \caption{Overlap of the (excited state) wave function of the realistic Coulombnic plus linear plus $A \text{exp}(- B r^{2})$ potential (represented by solid line) and that of the quadratic plus $A \text{exp}(-B r^{2})$ potential (represented by dashed line).}\label{1}
\end{center}
\end{figure}

Now we combine our Hamiltonian and all the wave functions we
have mentioned in the Schr$\ddot{\textrm{o}}$dinger equation for the meson-meson system, which means that the overlap of $(\textbf{H}-E_{c})\mid \Psi\rangle$
with an arbitrary variation $\mid\delta \Psi\rangle$ of state vector
$\mid \Psi\rangle$ vanishes where $\mid \Psi\rangle$ is the state
vector of the whole $q^{2}\overline{q}^{2}$ system. In $\mid \delta
\Psi\rangle$, we considered only the variation in $\chi_{k}$ (see
eq.\eqref{i7}), as in the resonating group method. Thus we wrote
\begin{align}
\begin{split}
\langle \delta \Psi\mid \textbf{H} - E_{c}& \mid \Psi\rangle = \sum_{k,l} \int d^{3}R_{c}d^{3}R_{K}d^{3}y_{K}d^{3}z_{K}\psi_{c}(\textbf{R}_{c})\delta
\chi_{k}(\textbf{R}_{K})\xi_{k}(\textbf{y}_{K})\xi_{k}(\textbf{z}_{K})\\& _{g}\langle k \mid \textbf{H}-E_{c}\mid  l\rangle_{g}\psi_{c}(\textbf{R}_{c})
\chi_{l}(\textbf{R}_{L})\xi_{l}(\textbf{y}_{L})\xi_{l}(\textbf{z}_{L}) = 0 \label{e4}
\end{split}
\end{align}
for$k,l=1, 2, 3, 1^{\star}, 2^{\star}, 3^{\star}$ and  $K, L = 1, 2,
3.$ The arbitrary variations $\delta \chi_{k}(\textbf{R}_{K})$'s for
different values of $R_{K}$ are linearly independent and hence their
co-efficient in eq.\eqref{e4} should be zero. With the trivial
$R_{c}$ integration performed to give a finite result, this leads to
\begin{align}
\sum_{l} \int d^{3}y_{K}d^{3}z_{K}\xi_{k}(\textbf{y}_{K})\xi_{k}(\textbf{z}_{K})_{g}\langle k \mid \textbf{H} - E_{c}\mid l
\rangle_{g}\chi_{l}(\textbf{R}_{L})\xi_{l}(\textbf{y}_{L})\xi_{l}(\textbf{z}_{L}) = 0, \label{e5}
\end{align}
where
\begin{align*}
_{g}\langle k \mid \textbf{H} - E_{c}\mid l
\rangle_{g} =_{g}\langle k \mid KE + V + 4 m - E_{c}\mid l
\rangle_{g}.
\end{align*}
Elements of $V$ and $KE$ matrices are defined below in eq.\eqref{e8}
and eq.\eqref{e9}. In ref.~\cite{P. Pennanen98} it is stated that,
in the lattice QCD simulations, it was found that the energy of the
lowest state was always the same in both a $2\times2$ and $3\times3$
description, provided $\mid 1 \rangle$ or $\mid 2 \rangle$ had the
lowest energy. In addition the energy of the second state was, in
most cases, more or less the same. Two level approximation is also
used in later works~\cite{Green99,V2005}
of the tetraquark system.
Considering this, we include only two topologies (1 and 2), meaning four
states$(\mid1\rangle, \mid2\rangle, \mid1^{\star}\rangle,
\mid2^{\star}\rangle)$. According to the model of ref.~\cite{P.
Pennanen98} the overlap matrix $\emph{N}$ in this truncated 4-basis
is given by
\begin{equation}
N =\{_{g}\langle k \mid l \rangle_{g}\} = \left[
\begin{array}{rrrr}
1 & f/3 & 0 & -f^{a}/3 \\
f/3 &  1 & -f^{a}/3 & 0 \\
0 & -f^{a}/3 & 1 & -f^{c}/3 \\
-f^{a}/3 & 0 & -f^{c}/3 & 1
\end{array} \right].\label{e6}
\end{equation}
And the potential matrix $V(f)$ is
\begin{equation}
V =\{_{g}\langle k\mid V \mid l \rangle_{g}\} = \left[
\begin{array}{rrrr}
V_{11} & V_{12} & V_{11^{\star}} & V_{12^{\star}} \\
V_{21} &  V_{22} & V_{21^{\star}} & V_{22^{\star}} \\
V_{1^{\star}1} & V_{1^{\star}2} & V_{1^{\star}1^{\star}} & V_{1^{\star}2^{\star}} \\
V_{2^{\star}1}& V_{2^{\star}2}& V_{2^{\star}1^{\star}} & V_{2^{\star}2^{\star}}
\end{array} \right].\label{e7}
\end{equation}
Here,
\begin{align}
\begin{split}
&V_{11} = -\frac{4}{3} (v_{1\overline{3}} + v_{2\overline{4}})\\
&V_{12} = V_{21} = \frac{4}{9} f(v_{12} + v_{\overline{3}\overline{4}} - v_{1\overline{3}} - v_{2\overline{4}} - v_{1\overline{4}} - v_{2\overline{3}})\\
&V_{22} = -\frac{4}{3} (v_{1\overline{4}} + v_{2\overline{3}})\\
&V_{21^{\star}} = V_{2^{\star}1} =
-\frac{f^{a}}{18}\bigg(\sqrt{2}(v^{\star}_{1\overline{3}} +
v^{\star}_{2\overline{4}}) -
\frac{16}{\sqrt{2}}(v^{\star}_{1\overline{4}} + v^{\star}_{2\overline{3}}) - \sqrt{2}(-v^{\star}_{12}-v^{\star}_{\overline{3}\overline{4}})\bigg)\\
&V_{1^{\star}2} = V_{12^{\star}} = -\frac{f^{a}}{18}\bigg(\sqrt{2}(v^{\star}_{1\overline{4}} + v^{\star}_{2\overline{3}}) - \frac{16}{\sqrt{2}}(v^{\star}_{1\overline{3}} + v^{\star}_{2\overline{4}}) - \sqrt{2}(-v^{\star}_{12} - v^{\star}_{\overline{3}\overline{4}})\bigg)\\
&V_{1^{\star}1^{\star}} = \frac{1}{6}( v^{\star}_{1\overline{3}} + v^{\star}_{2\overline{4}})\\
&V_{1^{\star}2^{\star}} = V_{2^{\star}1^{\star}} = -\frac{1}{18} f^{c}\bigg( -(v^{\star}_{1\overline{3}} + v^{\star}_{2\overline{4}} + v^{\star}_{1\overline{4}} + v^{\star}_{2\overline{3}}) + 10(v^{\star}_{12} + v^{\star}_{\overline{3}\overline{4}})\bigg)\\
&V_{2^{\star}2^{\star}}=\frac{1}{6}(v^{\star}_{1\overline{4}} + v^{\star}_{2\overline{3}})\\
&V_{1^{\star}1} = V_{11^{\star}} = V_{2^{\star}2} = V_{22^{\star}} =
0 , \label{e8}
\end{split}
\end{align}
with $v^{\star}_{ij} = v_{ij} + \epsilon \triangle v^{\star}_{ij}$, $\epsilon$ being defined above
(after eq.\eqref{i9}). The coefficients of $v_{ij}$ and $v^{\star}_{ij}$, resulting from the $\textbf{F}.\textbf{F}$ operator, are given in Table 1 in the Appendix A.
The kinetic energy matrix of the two quarks two anti-quarks is taken
to be
\begin{equation}
KE =\{_{g}\langle k\mid KE \mid l \rangle_{g}\} = \emph{N(f)}^{\frac{1}{2}}_{k,l}
\big(\frac{-1}{2
m}\sum^{4}_{i=1}\nabla^{2}_{i}\big)\emph{N(f)}^{\frac{1}{2}}_{k,l}. \label{e9}
\end{equation}
The kinetic energy in the same form is also used in
ref.~\cite{B. Masud91}.

The gluonic field overlap factor $\emph{f}$, as written in
introduction, is suggested by ref.~\cite{P. Pennanen98} as
\begin{equation}
f = \text{exp}[- b_{s} k_{f} S], \label{e10}
\end{equation}
with $b_{s} = 0.18GeV^{2}$~\cite{Isgur85}.
In ref.~\cite{Green99} the gluonic field overlap factor $f$ is used
in the Gaussian form as
\begin{equation}
f = \text{exp}[- k_{f} b_{s}\sum_{i<j}r^{2}_{ij}], \label{e11}
\end{equation}
employed in $SU(3)_{c}$ for interpreting the results in terms of the
potential for the corresponding single heavy-light meson. In
ref.~\cite{Green99}, the simulations that are fitted by using $f$
are for the configurations when the gluonic field is in the ground
state i.e. overlap matrix is a $2 \times 2$ matrix. In ref.~\cite{P.
Pennanen98}, simulations are reported with 2-colour approximation.
But in ref.~\cite{Green99}, lattice simulations performed for $SU(3)_{c}$
are reported. Our overlap, potential and kinetic energy matrices are
also written in $SU(3)_{c}$, so we use the $k_{f}$ multiplying sum
of area form of $f$ (written in eq.\eqref{e11}) with $k_{f}= 0.6$
(as used in ~\cite{Green99}) for numerical convenience and have
not used the minimal area form. When we are observing the dynamical effects
for the ground state, our overlap, potential and kinetic energy
matrices are $2 \times 2$ matrices and we use $f$ with $k_{f} =
0.6$. But when we incorporate the excited state gluonic field our
overlap, potential and kinetic energy matrices become $4 \times 4$
matrices, in the upper left $2 \times 2$ block of these matrices,
the form of $f$ remains the same but the value of $k_{f}$ is changed
to $1.51$ according to the conclusion of ref.~\cite{P. Pennanen98}.
In the other blocks $f^{a}$, $f^{c}$ are also used and defined in refs.~\cite{P. Pennanen98}~\cite{P. PennanenB} as
\begin{equation*}
f^a = (f_1^a + b_s f^a_2 S)\text{exp}[- b_{s} k_{a} S],
\end{equation*}
\begin{equation*}
f^c = \text{exp}[- b_{s} k_{c} S].
\end{equation*}
If we take
$f^{a}$ as a function of area as defined in ~\cite{P. Pennanen98},
it becomes unmanageable to solve the integral equations
(\ref{e12}-\ref{e15}) and hence we have taken $f^{a}$ to be a constant but have tried a
variety of its values to explore how much our conclusions depend on
its value. As for $f^{c}$, the fit in ref.~\cite{P. PennanenB}
~\cite{P. Pennanen98} of the model to the lattice data favours
$k_{c}= 0$ which implies that $f^{c} = 1$ i.e. the excited
configurations interact amongst themselves in the way expected from
perturbation theory. Thus we have used $f^{c} = 1$.
\begin{center}
\section*{3. COUPLED INTEGRAL EQUATIONS}
\end{center}
\qquad Using "\emph{N}","\emph{V}" and "$\emph{KE}$" elements in
eq.\eqref{e5}, we got four integral equations for four different
values of $\emph{k}$ or $\emph{l}$. Then we do $\textbf{y}_{K}$ and
$\textbf{z}_{K}$ integrations. All the integrations required are in the Gaussian form or modified Gaussian form (with a polynomial in the
integrand multiplying the Gaussian exponential) and we integrate analytically.
For $K=L=1,2$, in eq.\eqref{e5}, $\chi_{l}(\textbf{R}_{L})$ is
independent of $\textbf{y}_{K}$ and $\textbf{z}_{K}$ and, thus, can
be taken out of integrations. After the integration, the result is
$\textbf{R}_{K}$ dependent co-efficient of
$\chi_{k}(\textbf{R}_{K})$.
For $K \neq L$, $\textbf{y}_{K}$ and $\textbf{z}_{K}$ are replaced
by their linear combinations with one of them as identical to
$\textbf{R}_{L}$ and other one independent of it as
$\textbf{R}_{3}$. The jacobian of transformation is equal to 8. Then we integrate the
equation w.r.t $\textbf{R}_{3}$. Integration leaves the following four equations:
\begin{align}
\begin{split}
\Delta_{1}(\textbf{R}_{1})&\chi_{1}(\textbf{R}_{1}) +\int
d^{3}\textbf{R}_{2} F (\textbf{R}_1, \textbf{R}_2) \chi_{2}(\textbf{R}_{2}) +
\int d^{3}\textbf{R}_{2}E_1 (\textbf{R}_1, \textbf{R}_2)\chi^{\star}_{2}(\textbf{R}_{2}) = 0, \label{e12}
\end{split}
\end{align}
\begin{align}
\begin{split}
\Delta_{2}(\textbf{R}_{2}) &\chi_{2}(\textbf{R}_{2}) + \int
d^{3}\textbf{R}_{1} F (\textbf{R}_1, \textbf{R}_2) \chi_{1}(\textbf{R}_{1}) +
\int d^{3}\textbf{R}_{1} E_1 (\textbf{R}_2, \textbf{R}_1)\chi^{\star}_{1}(\textbf{R}_{1}) = 0 \label{e13}
\end{split}
\end{align}
\begin{align}
\begin{split}
F_{4}(\textbf{R}_{1})&\chi^{\star}_{1}(\textbf{R}_{1}) + \int
d^{3}\textbf{R}_{2} E_2 (\textbf{R}_1, \textbf{R}_2) \chi_{2}(\textbf{R}_{2}) + \int
d^{3}\textbf{R}_{2} E_3 (\textbf{R}_1, \textbf{R}_2)\chi^{\star}_{2}(\textbf{R}_{2}) = 0 \label{e14}
\end{split}
\end{align}
\begin{equation}
\begin{split}
F_{4}(\textbf{R}_{2})& \chi^{\star}_{2}(\textbf{R}_{2}) + \int
d^{3}\textbf{R}_{1} E_2 (\textbf{R}_2, \textbf{R}_1) \chi_{1}(\textbf{R}_{1}) + \int
d^{3}\textbf{R}_{1} E_3 (\textbf{R}_2, \textbf{R}_1) \chi^{\star}_{1}(\textbf{R}_{1}) = 0. \label{e15}
\end{split}
\end{equation}
The symbols are defined in the appendix B. We have eventually
replaced $\textbf{r}_{1}, \textbf{r}_{2}, \textbf{r}_{3},
\textbf{r}_{4}$ by $\textbf{R}_{1}, \textbf{R}_{2}, \textbf{R}_{3}$,
and $\textbf{R}_{c}$. With trivial integration on $\textbf{R}_{c}$,
we have eq.\eqref{e5} that is independent of $\textbf{R}_{c}$. Now,
after the integration on $\textbf{R}_{3}$, the above four integral
equations (\ref{e12}-\ref{e15}) depend only on $\textbf{R}_{1}$ and
$\textbf{R}_{2}$. So every quantity which we want to calculate
depends on $\textbf{R}_{1}$ and $\textbf{R}_{2}$.
In eq.(\ref{e12}-\ref{e13}), the first two terms  in each equation,
containing $\chi_{1}(\textbf{R}_{1})$ and $\chi_{2}(\textbf{R}_{2})$,
are for the ground state. It is noted that in these terms (observing the definitions of the symbols in appendix B), there is no dot product of vectors
$\textbf{R}_{1}$ and $\textbf{R}_{2}$. So the results from these
terms should not depend on the angle between $\textbf{R}_{1}$ and
$\textbf{R}_{2}$, called $\theta$. The third term
in each of eq.(\ref{e12}-\ref{e13})
is due to the gluonic ground and excited states. In these terms dot
product of two vectors ($\textbf{R}_{1}$ and $\textbf{R}_{2}$)
appear, so the results from these terms depend on $\theta$.

\section*{\begin{center} 4. SOLVING THE INTEGRAL EQUATIONS \end{center}}
\qquad Now taking the three dimensional Fourier transform of
eq.(\ref{e12},\ref{e14}) with respect to $\textbf{R}_{1}$ and
eq.(\ref{e13},\ref{e15}) with respect to $\textbf{R}_{2}$, we get
formal solutions $\chi_{1}(\textbf{P}_{1})$,
$\chi_{2}(\textbf{P}_{2})$, $\chi^{\star}_{1}(\textbf{P}_{1s})$, and
$\chi^{\star}_{1}(\textbf{P}_{1s})$ as shown in appendix C. Because
of the coupling to the gluonic excitations, it become difficult to
solve the integral equations for non trivial solutions for
$\chi_{1}(\textbf{P}_{1}), \chi_{2}(\textbf{P}_{2}),
\chi^{\star}_{1}(\textbf{P}_{1s})$, and $
\chi^{\star}_{2}(\textbf{P}_{2s})$ analytically as done in ~\cite{B.
Masud91,B. Masud94}. In ~\cite{B. Masud91}, the meson wave
functions, including the gluonic field overlap factor, is
$\textbf{R}_{1}$, $\textbf{R}_{2}$ separable. So there the integral
equations can be solved analytically by replacing $\chi_{1}$ and
$\chi_{2}$. But in our present work, the meson-meson wave functions
are not separable in $\textbf{R}_{1}$,$\textbf{R}_{2}$. So we use
the Born approximation (as used in ~\cite{T. Barnes92} for meson-meson scattering) to solve the integral equations. Our results given
below also justify our use of the Born Approximation.
Using this approximation, we use the  solutions
($\chi_{i}(\textbf{R}_{i}),\chi^{\star}_{i}(\textbf{R}_{i})$) of
eqs.(\ref{e12}-\ref{e15}) in absence of interactions (meaning $f =
f^{a} = f^{c} = 0$)
\begin{equation}
\begin{split} &\chi_{i}(\textbf{R}_{i}) = \sqrt{\frac{2}{\pi}}\quad \text{exp}(\imath
\textbf{P}_{i}.\textbf{R}_{i}),\\ \textrm{and}\qquad \qquad
&\chi^{\star}_{i}(\textbf{R}_{i}) = \sqrt{\frac{2}{\pi}}\quad \text{exp}(\imath
\textbf{P}_{is}.\textbf{R}_{i}) \label{e16}
\end{split}
\end{equation}
for $i = 1, 2$. Here the coefficient of $\text{exp}(\imath
\textbf{P}_{i}.\textbf{R}_{i})$ is chosen so that it makes
$\chi(\textbf{R}_{i})$ as Fourier transform of
$\frac{\delta(P_{i}-P_{c}(i))}{P^{2}_{c}(i)}$. Similarly the
coefficient of $\text{exp}(\imath \textbf{P}_{is}.\textbf{R}_{i})$
is chosen. Using this approximation, the integration on
$\textbf{R}_{1}$ and $\textbf{R}_{2}$ can be performed to get
$\chi_{1}(\textbf{P}_{1})$ (written in eq.(C11)).

 $T_{11}$ can be calculated (as in
ref.~\cite{B. Masud91}) by considering the coefficient of
$\frac{1}{\triangle_{1}(P_{1})}$ containing the
$\chi_{1}(\textbf{R}_{1})$ from eq.(C11). As in this equation, there
is no coefficient having $\chi_{1}(\textbf{R}_{1})$, so it gives
$T_{11}=0$. $T_{12}$ can be calculated by considering the
coefficient of $\frac{1}{\triangle_{1}(P_{1})}$ containing the
$\chi_{2}(\textbf{R}_{2})$ from eq.(C11) in the following eq.
\begin{align}
T_{12}= M \frac{\pi}{2} P_{1} [\textrm{coef.}\quad \textrm{of} \quad \frac{1}{\triangle_{1}(P_{1})}\quad \textrm{containing} \quad \chi_{2}(\textbf{R}_{2})], \label{e17}
\end{align}
with $M$ being the mass of $c\overline{c}$ meson. Similarly
$T_{12^{\star}}$ can be calculated by considering the coefficient of
$\frac{1}{\triangle_{1}(P_{1})}$ containing the
$\chi^{\star}_{2}(\textbf{R}_{2})$ from eq.(C11)
\begin{align}
T_{12^{\star}}= M \frac{\pi}{2} P_{1s} [\textrm{coef.}\quad \textrm{of} \frac{1}{\triangle_{1}(P_{1})}\quad \textrm{containing} \quad
\chi^{\star}_{2}(\textbf{R}_{2})]. \label{e18}
\end{align}
The relation between off-diagonal transition and scattering matrix element is written as
\begin{align}
S=I- 2 \iota T, \label{e19}
\end{align}
where $S$, $T$, and $I$ represent $4 \times 4$ scattering,
transition, and identity matrices respectively.
The eq.\eqref{e19} can also be written as
\begin{align}
S_{ij}=\delta_{ij}- 2 \iota T_{ij}. \label{e20}
\end{align}
for $i,j=1,2,1^{\star},2^{\star}$.
Using the transition matrix elements, the contribution to the energy shift of meson-meson system ($cc\overline{c}\overline{c}$) through $\ell=0$ states (both with and without gluonic excitations) can be calculated by
using the stationary state perturbation theory, i.e.
\begin{align}
E_{i} = E^{0}_{i} + T_{ii}+\sum_{i\neq
j}\int^{\infty}_{0}\frac{|T_{ij}|^{2}}{E^{0}_{i}-E^{0}_{j}}dP_{j}, \label{e21}
\end{align}
with the initial state $i$ and intermediate state $j$.
We have considered initial states where the gluonic field should is in ground state, so $i=1,2$,
but intermediate $j=1,2,1^{\star},2^{\star}$. Here $T_{ii} = \langle
i|T|i\rangle$, $|T_{ij}|^{2} = |\langle j|T|i\rangle|^{2}$,
$E^{0}_{i}$ is the energy of a ground state (1 or 2) of meson-meson
system, and $E^{0}_{j}$ may be the energy of the other meson-meson
ground state or that of an gluonic-excited meson-meson state.
\section*{\begin{center}5. RESULTS AND CONCLUSIONS \end{center} }

\qquad 1-The transition amplitude $T_{12}$, from one meson-meson ground
state to other, is calculated by using eq.\eqref{e17} with
$k_{f}=0.6$~\cite{Green99}(without the incorporation of gluonic
excited states). Its dependence on the center of mass kinetic energy
is shown in Fig. 4 below. As it is noted that magnitude of transition element $T_{12}$
is less than 1, so this result shows the validity of Born
approximation. In result 3 we compare the transition amplitude of this many
body ground state gluonic field model at $k_{f}=0.6$ with the transition amplitude obtained from a model that is extended to gluonic excitations
along with changing $k_{f} = 1.51$.

2- The transition matrix element $T_{12^{\star}}$, for transition from
ground state to excited state gluonic field with $k_{f}=1.51$,
depend on the parameters $\epsilon$, and $f^{a}$. $T_{12^{\star}}$
also depends on $\theta$ (the angle that $\textbf{P}_{1}$ makes with
$\textbf{P}_{2}$ and $\textbf{P}_{1s}$ makes with
$\textbf{P}_{2s}$). We take parameter $f^{a}$ as a constant as
discussed earlier in section 2. For $\epsilon = 2$, we
take different values of $f^{a}$ to see the effects of $f^{a}$ on
$T_{12^{\star}}$. Fig. 5 shows the dependence of
$T_{12^{\star}}$ on $f^{a}$ at $\theta = 90$. By taking different values of $\theta$, $T_{12^{\star}}$ was calculated. We found that the results are slightly different for different $\theta$.
This slight angle dependence is not directly reported here, but can be found by a linear combination of the corresponding $m=0$ spherical harmonics with coefficients for each value of energy read from Figs. (8-10) that report the partial wave amplitudes that result from this angle dependence. For $\epsilon = 1/2$, the dependence of
$T^{\star}_{12}$ on $f^{a}$ is shown in Fig. 6. And
for $\epsilon = 1$, the dependence of $T^{\star}_{12}$ on $f^{a}$ is shown in Fig. 7. These graphs show that the magnitude of transition amplitude is increasing with the increase
of $f^{a}$.
\begin{figure}
\begin{center}
\epsfig{file=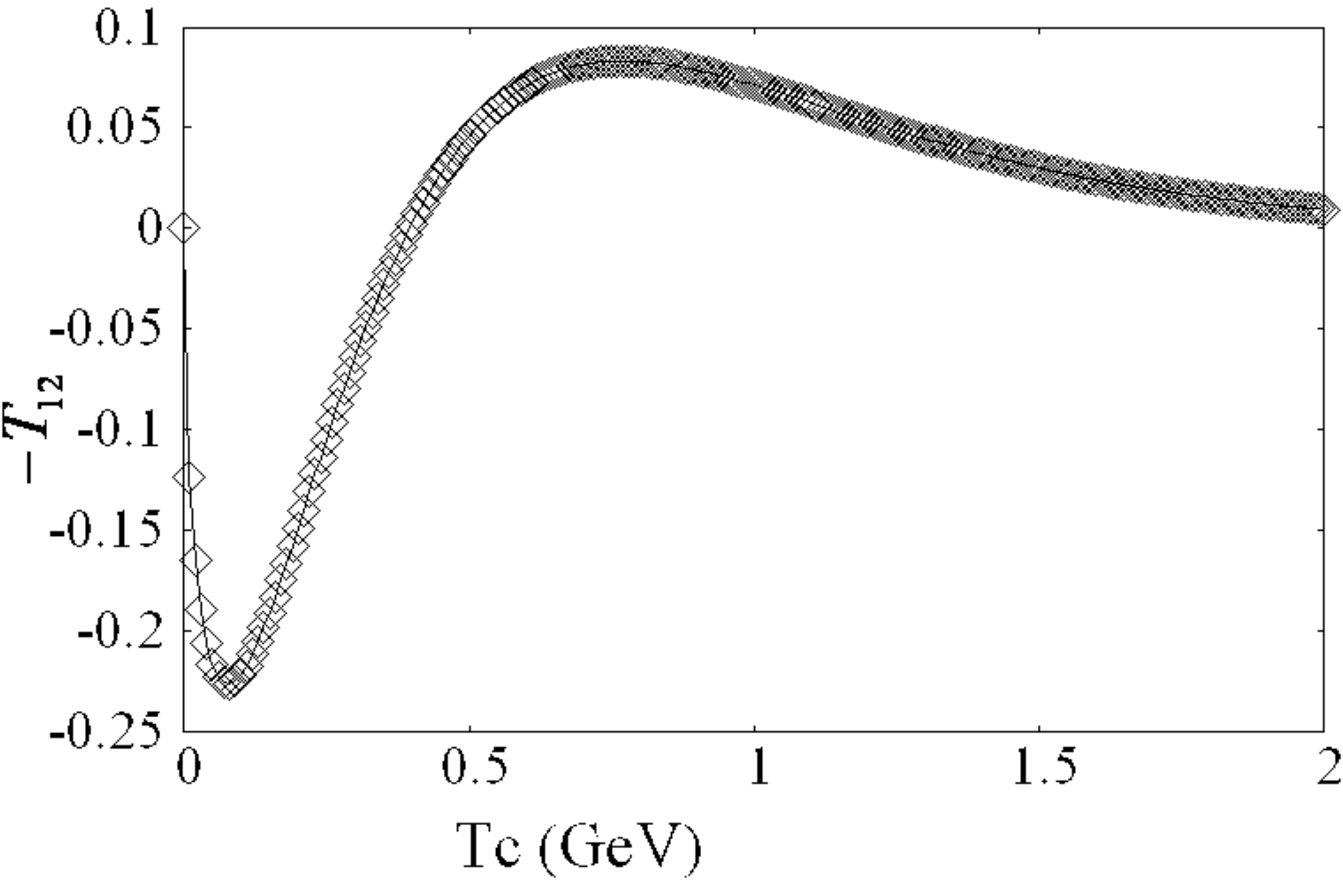,width=0.9\linewidth,clip=}\caption{The graph for the energy \emph{vs.} $T_{12}$ for ground state at $k_{f} = 0.6.$}
\end{center}
\end{figure}
\begin{figure}
\begin{center}
\epsfig{file=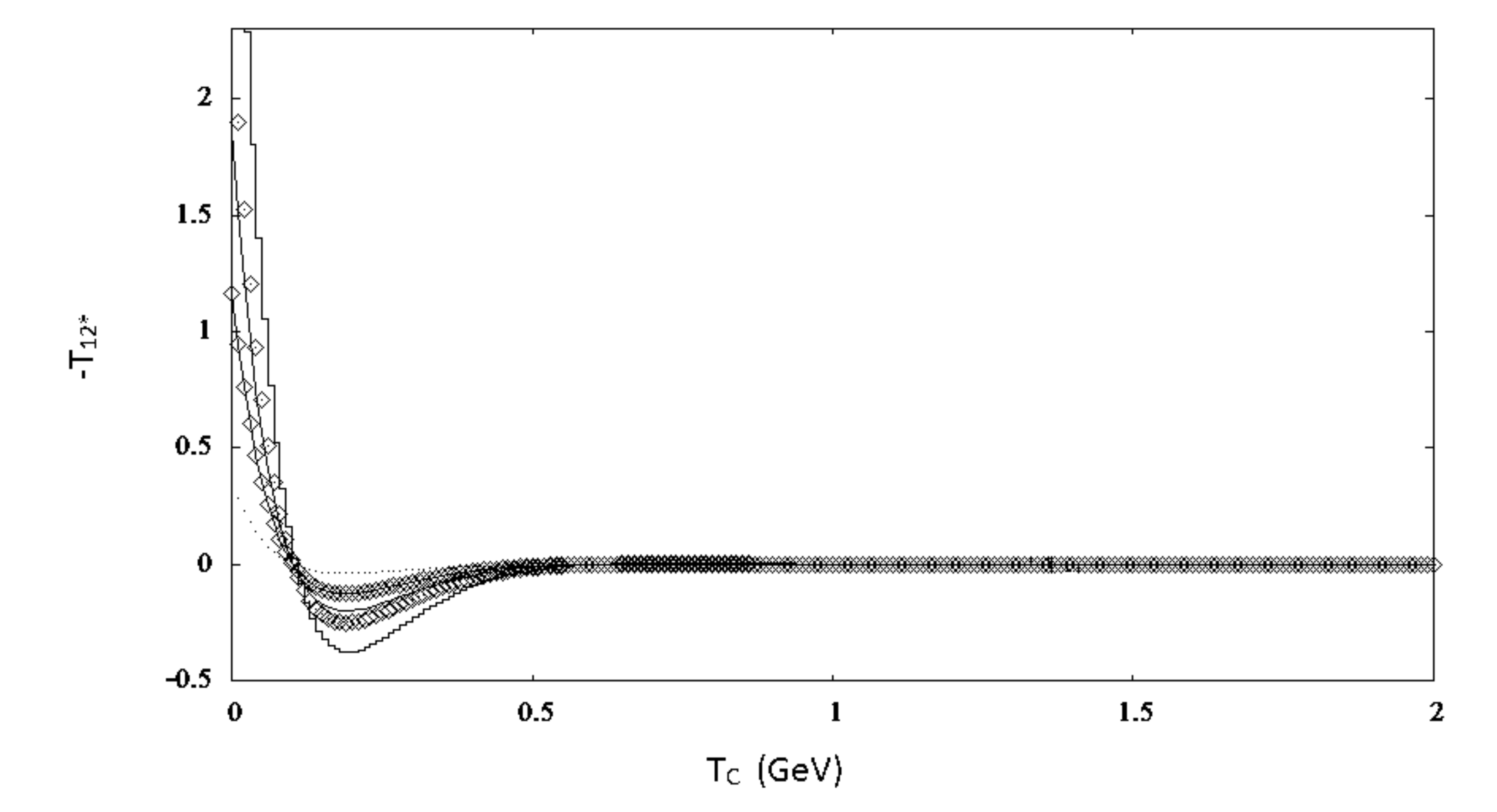,width=1\linewidth,clip=}\caption{The graph for the energy \emph{vs.} $T_{12^{\star}}$ at $\theta = 90$ with $\epsilon=2$ for different
values of $f^{a}$. The curve with dots is for $f^{a}=0.015$, with lines plus points is for $f^{a}= 0.05$, with lines is for $f^{a}=0.08$, with points
is for $f^{a}=0.1$, and with steps is for $f^{a}=0.15$.}
\end{center}
\end{figure}
\begin{figure}
\begin{center}
\epsfig{file=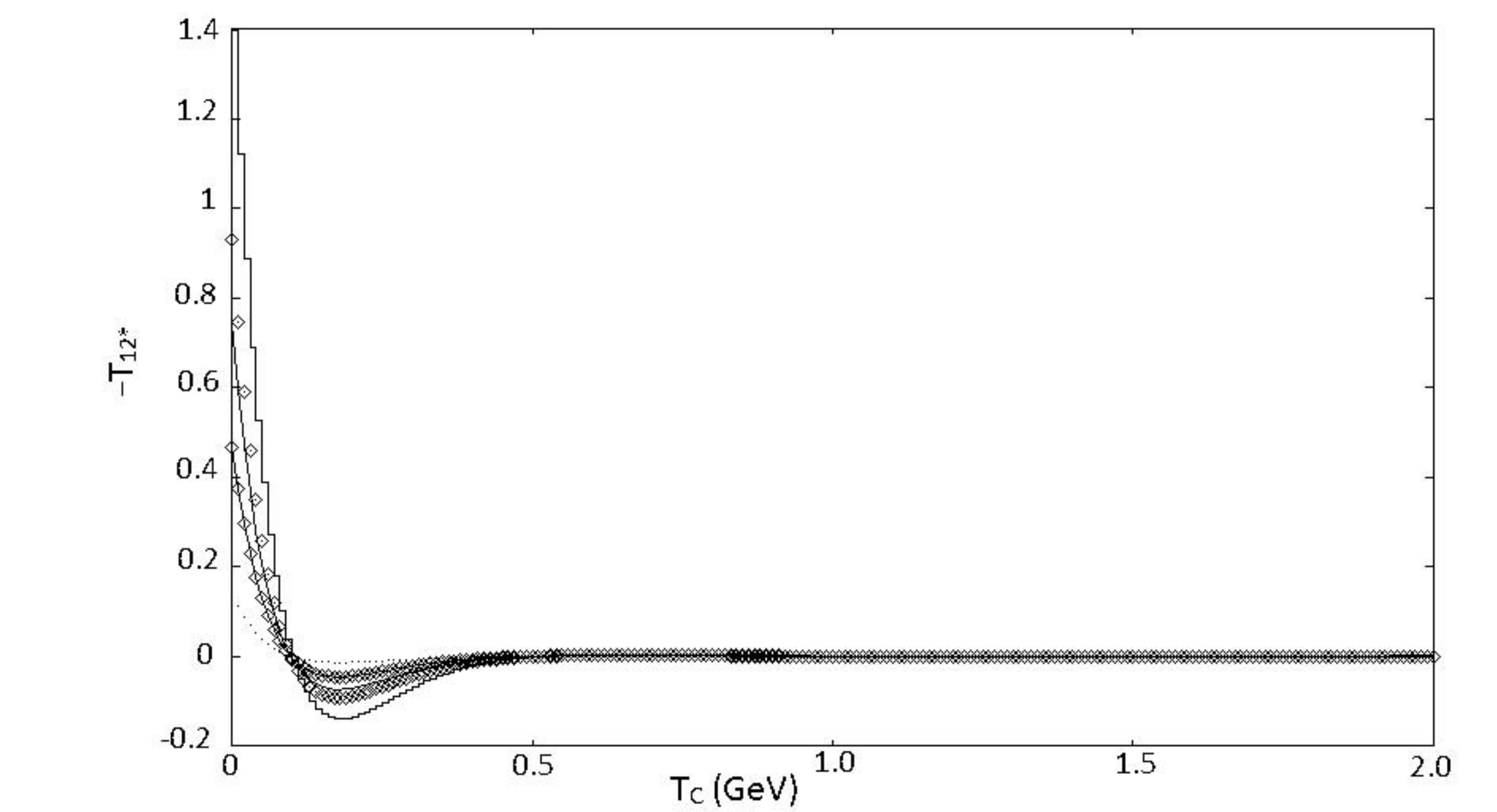,width=1\linewidth,clip=}\caption{The same as Fig.\ 5 but with $\epsilon=1/2$.}
\end{center}
\end{figure}
\begin{figure}
\begin{center}
\epsfig{file=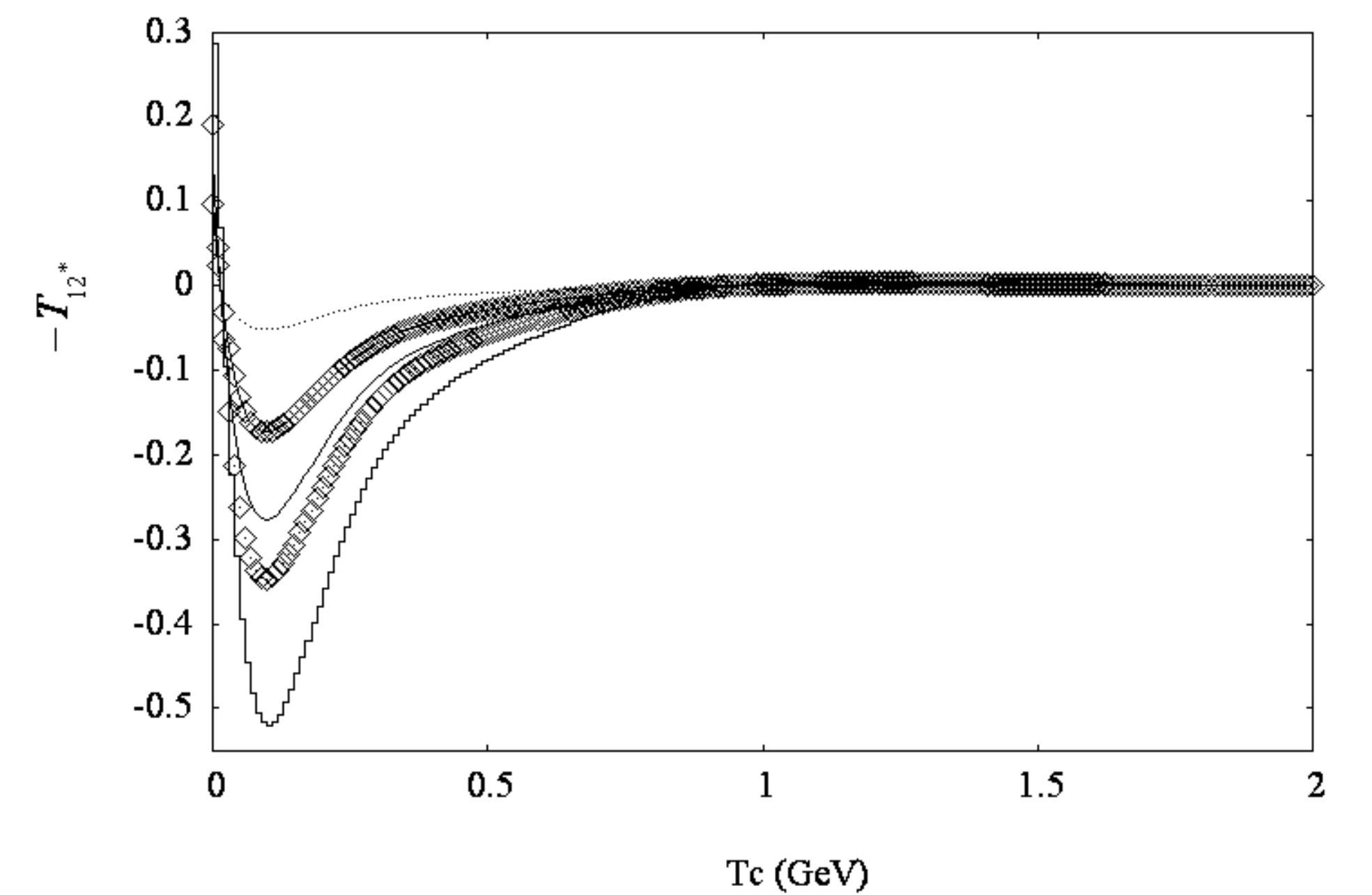,width=1\linewidth,clip=}\caption{The same as Fig.\ 5 but with $\epsilon=1$.}
\end{center}
\end{figure}

3- For the center of mass kinetic energy range $0$ to $2$ GeV, the
average of the modulus of transition amplitudes (excluding excited states)
is equal to $0.0369$ at $k_{f}=0.6$.
But if we change the value of $k_{f}$ from 0.6 to 1.51 and also
include the excited state gluonic field, the average meson-meson
transition amplitude again becomes $(0.0218 + 0.0152)$ =
0.0370 by choosing $f^{a}=0.03$ and $\theta = 90$ with $\epsilon
= 2$. This means that a sum-of-two-body potential model can replace
to some extent many-body potential terms in a tetraquark system by
including the gluonic excitations. (An exact sum of two body terms would required $f_a=1$ though). In above, $0.0218$ is the transition amplitude due to ground state gluonic field at $k_{f}= 1.51$, and $0.0152$ is the transition amplitude due to gluonic excited states.

We have also explored the results with $\epsilon=1/2$ and
$\epsilon=1$. As $\epsilon $ is proportional to the potential matrix
elements taken between ground and gluonic excited states, when we
replace $\epsilon=2$ with $\epsilon=1/2 \quad \text {or}\quad 1$ ,
it has no effect on $T_{12}$ . But for
$\epsilon=\frac{1}{2}$ with inclusion of gluonic excitations, now the values $f^{a}=0.08$
and $\theta = 75$ almost restore the average transition amplitudes for
$k_{f}=1.51$ to $(0.0218 + 0.0149)$  = 0.0367 . (The significance of
the first and second term is as before). For $\epsilon=1$, at
$f^{a}=0.055$ and $\theta = 90$, the average transition amplitudes for $k_{f}
=1.51$ becomes $(0.0218 + 0.0150)$  =
0.0368. This almost restoration again indicates that
perhaps we can always choose parameters etc. so that with an
inclusion of the gluonic excitations the sum-of-two-body potential
model can replace many-body potential terms in a tetraquark system.

 4- The effects of interaction are also observed on a contribution to the
 energy shift by using eq.\eqref{e21}.
Here $T_{ii}=0$
with $i=1,2$. In $\int_{i\neq
j}\frac{|T_{ij}|^{2}}{E^{0}_{i}-E^{0}_{j}}dP$, we have chosen $E^{0}_{i} = 5.9176$ GeV, the value of energy for which center of mass kinetic energy becomes zero.  The intermediate energy state $E^{0}_{j}$ we study
depends on the state of gluonic field \big( $|2\rangle,
|2^{\star}\rangle$ are only possible states that couple to
$|1\rangle$\big). For the gluonic ground state ($|2\rangle$), $E^{0}_{j}= 5.9176 +
0.3380 P_{1}^{2}$ (calculated using eq.(C10) with $P^{2}_{c}(1) =
P^{2}_{1}$), and for the excited state gluonic field
($|2^{\star}\rangle$), $E^{0}_{j} = 5.4638 + 0.3380 P^{2}_{1s}$
(calculated using eq.(C10) with $P^{2}_{c}(1s) = P^{2}_{1s}$). At
$k_{f}=0.6$ and $E_{c} = 5.9176$ (i.e. at the threshold), the shift
to the ground state meson-meson energy is found to be
$E_{i}=E^{0}_{i}-0.7268$ GeV excluding the gluonic excitations. With increasing  $k_{f}$ to 1.51 as usual and including the coupling
to the  gluonic-excited meson-meson state, we want to restore the
same energy shifts we can get to (with same origin of the first and
second correction terms)
\begin{align*}
E_{i}=E^{0}_{i}+(-0.0950 - 0.6371) \textrm{GeV} = E^{0}_{i} -0.7321 \textrm{GeV}
\end{align*}
if $f^{a}= 0.14$ is used. This shows that gluonic excitations can
replace the many terms for $\ell=0$
energy shifts as well. We note that the
energy shift is independent of $\theta$, the angle between
$\textbf{R}_{1}$ and $\textbf{R}_{2}$.

The energy shifts we have
reported here can be compared with the hadron-loops-generated mass
shifts to charmonium states reported in Table $\textrm{III}$ of ref.~\cite{Barnes08}.
It is difficult, though, to conclude anything from this comparison
as in ref.~\cite{Barnes08} the integrand contains squares of the
matrix elements of the $^{3}P_{0}$ meson decay amplitudes whereas in
our integrands in eq.\eqref{e21} contain squares of meson-meson
couplings. Thus though the intermediate states in both works are
respective \emph{hadron loops} $q\bar{q}q\bar{q}$, the initial and
final states in ref.~\cite{Barnes08} are $q\bar{q}$ but in our work
initial and final states are also $q\bar{q}q\bar{q}$. Only we
include the glounic-exited intermediate $q\bar{q}q\bar{q}$ states
(i.e. the hybrid hadronic loops) for our problem.

In result 3, it is
noted that the average transition amplitude obtained (for $\epsilon = 2$) by
a model that does not include the gluonic excitations is equal to
the average transition amplitude obtained by including gluonic excitations
for $f^{a}=0.03$, but the energy shift obtained by both models
becomes comparable at $f^{a}=0.14$. One possible reason of this
difference in the values of $f^{a}$ could be that the average transition amplitude is calculated for the center of mass kinetic energy range in
between 0 GeV to 2 GeV, but energy shift is calculated at threshold
center of mass kinetic energy.

5- For the ground state gluonic field, transition matrix elements from state 1 to 2 and from 2 to 1
are $\theta$ independent. But for transition elements
to the gluonic-excited meson-meson state depends on
$\theta$. We projected this angle dependence on spherical harmonics $Y_{lm}(\theta)$.
The results of this partial wave analysis are also reported in Figs. (8-10) for $m=0$ and $f^{a}=0.03$ (this value is used above in
result 3). The reason for truncating the spherical expansion to $m=0$ harmonics is that we have no dependence on $\phi$, the azimuthal. These analysis shows that partial wave
amplitudes are decreasing as we go from $Y_{00}$ the coefficient to the $Y_{50}$ coefficient i.e. from S-wave to H-wave.  Figs. 8 and 9 are for the even wave ($D,G,...$) ratios with
$S-$wave. S/D ratios are also used in ref.~\cite{Swanson94,T.
Barnes96}. These graphs shows that the $S/G$ ratio is too much large as
compared to the $S/D$ ratio. It means that $S-$wave is dominant over
$G-$wave. In Fig. 10, ratios of odd waves with $S-$wave are shown. We noted that $S/H$ is too
much large as compared to $S/P$. This shows that $H,J,...$ waves can
be neglected as compared to $S-$wave. The partial wave analysis
indicates the presence of $P,D,F,G,H$ waves only when we include the
gluonic excitations in combination with essentially sum-of-pair-wise
approach. It means that, in the presence of gluonic excitations, an
$\ell = 0,1,2,3,...$ ground state meson-meson system may couple to
$\ell = 0,1,2,3,...$ hybrid-hybrid systems as a intermediate states
or as final states.
\begin{figure}
\begin{center}
\epsfig{file=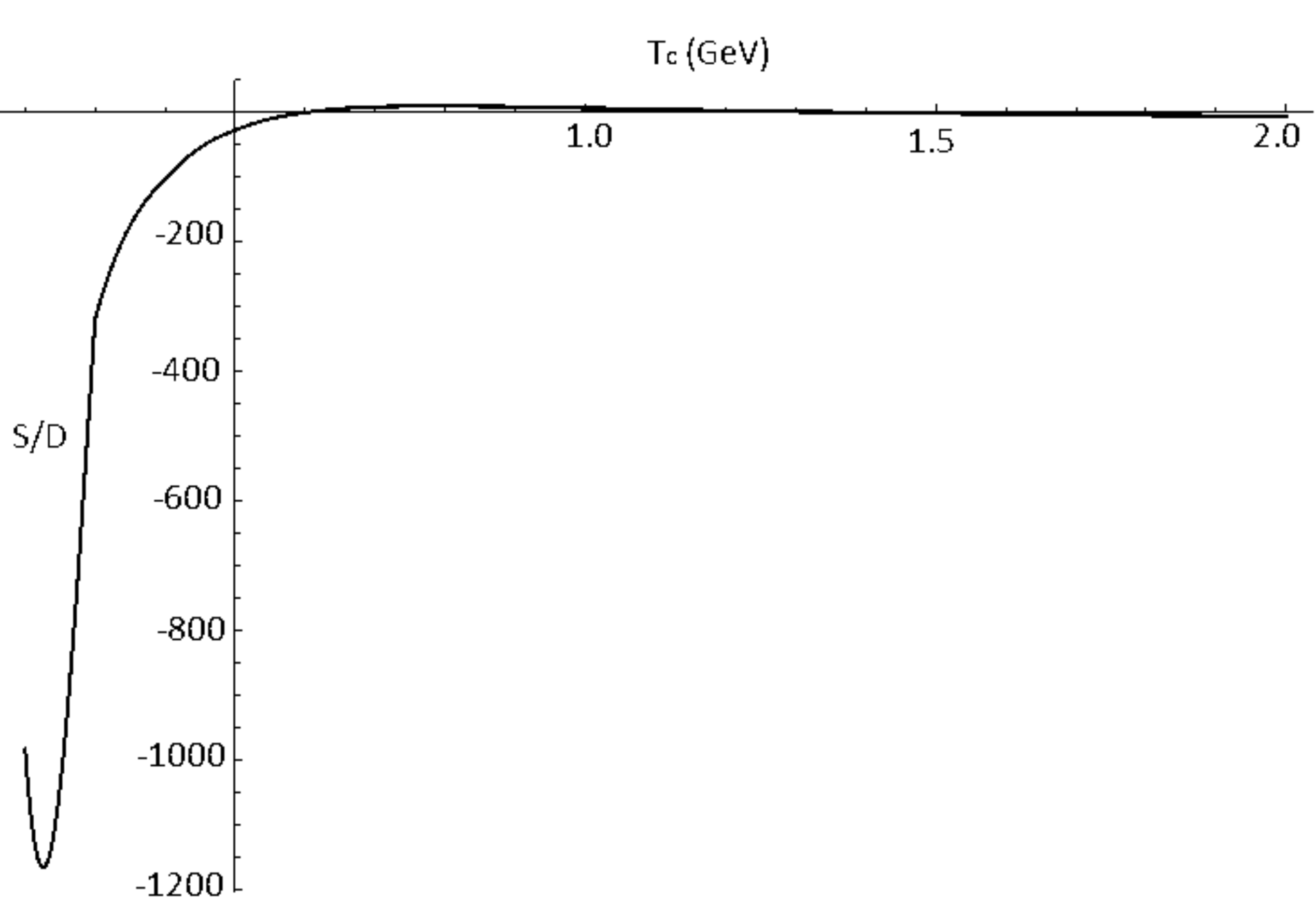,width=0.9\linewidth,clip=}\caption{ $T_{c}$ \emph{vs.} $S/D$ ratio.}
\end{center}
\end{figure}
\begin{figure}
\begin{center}
\epsfig{file=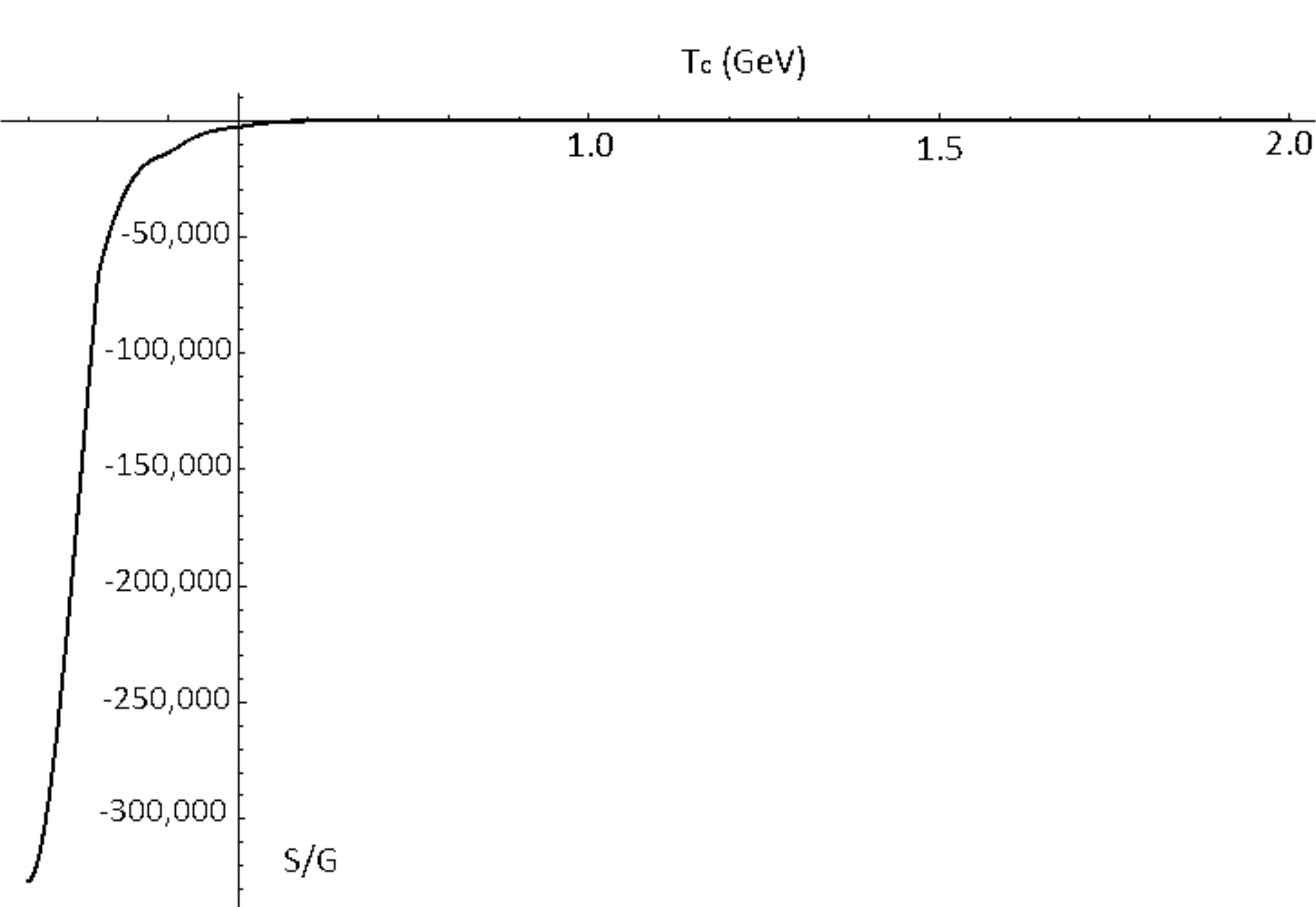,width=0.9\linewidth,clip=}\caption{ $T_{c}$ \emph{vs.} $S/G$ ratio.}
\end{center}
\end{figure}
\begin{figure}
\begin{center}
\epsfig{file=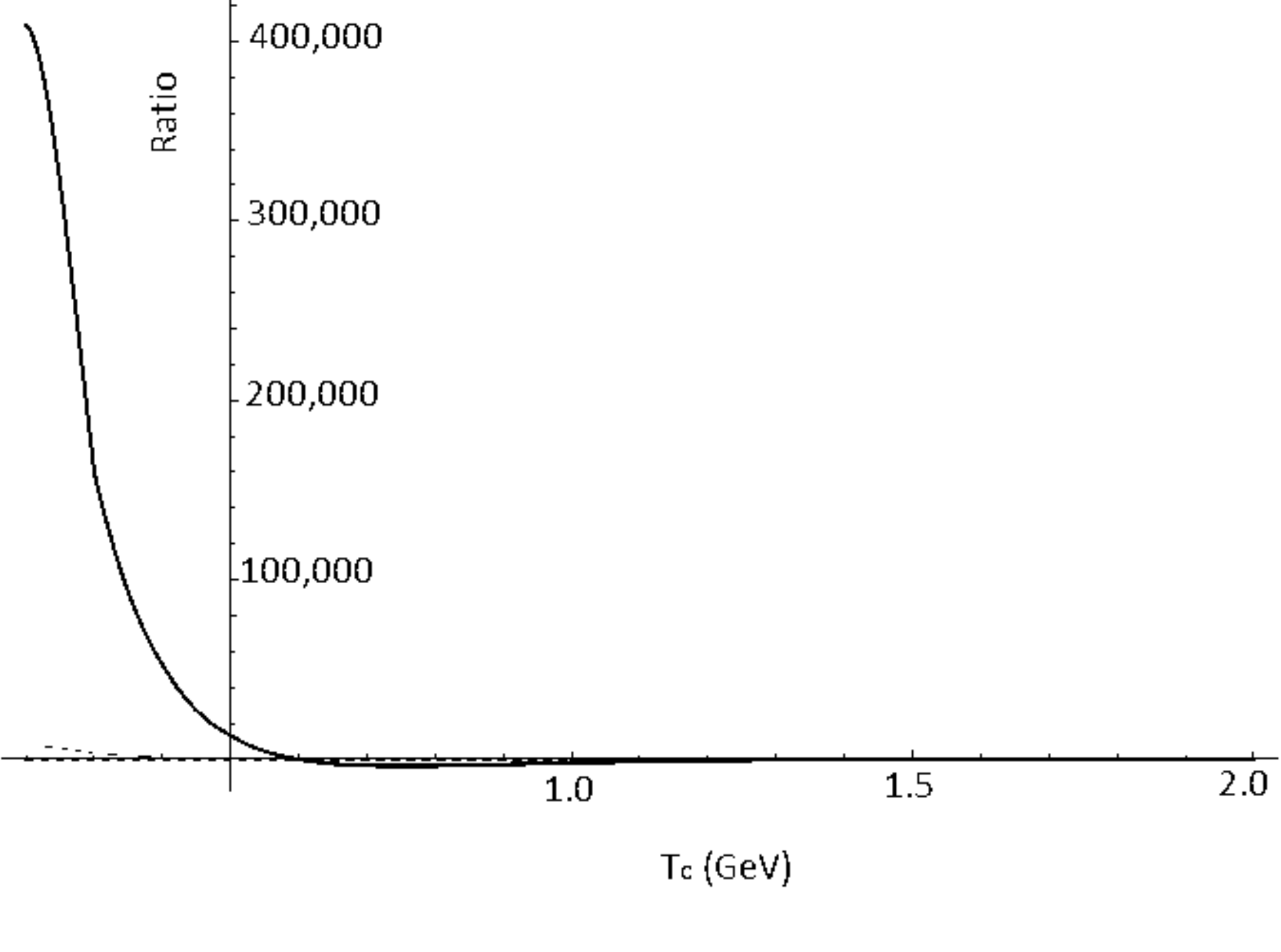,width=0.9\linewidth,clip=}\caption{ $T_{c}$ \emph{vs.} $S/P$ (Dashed line), $S/F$ (thin solid line), $S/H$ (thick solid line) ratios. $S/P$ and $S/F$ are so small as compared to $S/H$ that the curves for $S/P$ and $S/F$ are lie along horizontal axis.}
\end{center}
\end{figure}

As mentioned above in result 3,4, we compare our meson-meson transition amplitudes and
polarization potentials (in the respective center of mass energy
ranges) in an extended almost sum-of-pair-wise approach and a
many-body-term based one. Apparently this comparison has even a
qualitative difference in case of the transition amplitudes
($T_{12}$ and $T^{\star}_{12}$) which have
a dependance on the scattering angle (meaning coupling to $P$ and
higher orbital excitations in the final channel) if we use gluonic
excitation in combination with essentially sum-of-pair-wise approach
(can see Fig. 5)
but we get the same transition amplitude for all the scattering
angles if we use a many-body potential linking ground-state clusters
only. But a recent work \cite{Imran} indicates that the absence of
angle dependence is an artifact of the unjustified overly simple
form of the gluonic overlap factor $f$ (defined in eq.\eqref{e11}); using a proper lattice-gauge theory-based form of $f$ (defined in
eq.\eqref{e10})
also results in this
angle dependence and the resulting coupling to higher orbital
excitations.

The meson-meson to hybrid-hybrid coupling worked out in this paper may
affect properties of any system that is presently understood to be a
purely ground-state meson-meson system. This highlights, in addition
the energy shift of the meson-meson molecules due to coupling to
hybrid-hybrid systems, pointing out the possibility of a
hybrid-hybrid component in the wave functions of mesons like
$X(3872)$, $X(3940)$, $Y(4260)$, and $Z(4433)$ which are considered
to  have $q^{2}\overline{q}^{2}$ components. And in the above
paragraph, we indicate that this coupling may also include coupling
to orbital excitations. Thus we hope to point out a new dimension in discussions
about the structure of some exotic
 mesons as well.
\section*{\begin{center}ACKNOWLEDGEMENT\end{center}}
\qquad We are grateful to Higher education Commission of Pakistan for their
financial support no.17-5-3 (Ps3-212) HEC/Sch/2006.
\begin{center}
\section*{APPENDIX A}
\end{center}
\qquad Table 1 is used for the matrix elements of the $F_{i}.F_{j}$
operators (introduced above in eq.\eqref{i9}), for different values
of indices $i$ and $j$. In this table ground state elements (with
out star) are taken from ~\cite{B. Masud91} and excited state
elements (with star) are calculated by using the following group
theory results:
$$|8_{1 \overline{3}} 8_{2 \overline{4}}> = -\sqrt{\frac{2}{3}}|\overline{3}_{1 2} 3_{\overline{3} \overline{4}}> + \sqrt{\frac{1}{3}}|6_{1 2} \overline{6}_{3 4}>$$
$$|1_{1 \overline{3}} 1_{2 \overline{4}}> = \sqrt{\frac{1}{3}}|\overline{3}_{1 2} 3_{\overline{3} \overline{4}}> + \sqrt{\frac{2}{3}}|6_{1 2} \overline{6}_{\overline{3} \overline{4}}>$$
$$|8_{1 \overline{4}} 8_{2 \overline{3}}> = \sqrt{\frac{8}{9}}|1_{1 \overline{3}} 1_{2 \overline{4}}> - \sqrt{\frac{1}{9}}|8_{1 \overline{3}} 8_{2
\overline{4}}>$$
$$|1_{1 \overline{4}} 1_{2 \overline{3}}> = \sqrt{\frac{1}{9}}|\overline{1}_{1 \overline{3}} 1_{2 \overline{4}}> + \sqrt{\frac{8}{9}}|8_{1 \overline{3}} 8_{2
\overline{4}}>$$
$$= -\sqrt{\frac{1}{3}}|\overline{3}_{1 2} 3_{\overline{3} \overline{4}}> + \sqrt{\frac{2}{3}}|6_{1 2} \overline{6}_{\overline{3} \overline{4}}>$$
and
$$<8_{1 \overline{3}} 8_{2 \overline{4}}| \textbf{F}_{i}. \textbf{F}_{j}|8_{1 \overline{3}} 8_{2 \overline{4}}> = \frac{1}{6}.$$
\begin{table}\caption{The matrix elements of the $F_{i}.F_{j}$ operators}
\begin{tabular}{|c|c|c|c|c|c|c|c|c|}
\hline
& $<1| ..|1>$&$<1|..|2>$&$<2|..|2>$&$<1|..|2^{\star}>$&$<2|..|1^{\star}>$&$<1^{\star}|..|2^{\star}>$&$<1^{\star}|..|1^{\star}>$&$<2^{\star}|..|2^{\star}$\\
\hline
$ F_{1}.F_{2} $ &  0 & $\frac{4}{9}$ & 0 & $-\frac{2}{9\sqrt{2}}$ & $-\frac{2}{9\sqrt{2}}$ &$\frac{10}{18}$ & $-\frac{1}{3}$ &$-\frac{1}{3}$\\
$ F_{1}.F_{3}$ & $ -\frac{4}{3}$ & $-\frac{4}{9}$ & $0$ & $-\frac{16}{9\sqrt{2}}$ & $\frac{2}{9\sqrt{2}}$& $-\frac{1}{18}$ &$\frac{1}{6}$ & $-\frac{7}{6}$\\
$ F_{1}.F_{4}$&  0 & $-\frac{4}{9}$ & $-\frac{4}{3}$ & $\frac{2}{9\sqrt{2}}$ & $-\frac{16}{9\sqrt{2}}$ &$-\frac{1}{18}$ & $-\frac{7}{6}$ &$\frac{1}{6}$\\
$ F_{2}.F_{3}$&  0 & $-\frac{4}{9}$ & $-\frac{4}{3}$ & $\frac{2}{9\sqrt{2}}$ & $-\frac{16}{9\sqrt{2}}$ &$-\frac{1}{18}$ & $-\frac{7}{6}$ &$\frac{1}{6}$\\
$ F_{2}.F_{4}$&$ -\frac{4}{3}$ & $-\frac{4}{9}$ & $0$ & $-\frac{16}{9\sqrt{2}}$ & $\frac{2}{9\sqrt{2}}$& $-\frac{1}{18}$ &$\frac{1}{6}$ & $-\frac{7}{6}$\\
$ F_{3}.F_{4}$&  0 & $\frac{4}{9}$ & 0 & $-\frac{2}{9\sqrt{2}}$ & $-\frac{2}{9\sqrt{2}}$ &$\frac{10}{18}$ & $-\frac{1}{3}$ &$-\frac{1}{3}$\\
\hline
\end{tabular}
\end{table}

\section*{APPENDIX B}
\qquad The terms used in eqs.(\ref{e12}-\ref{e15}) are defined as
\begin{align*}
\Delta_{i}(\textbf{R}_{i}) = -E_{c}- 8 C d^{2}-\frac{8}{3}\overline{C}-\frac{\nabla_{\textbf{R}_{i}}^{2}}{2 m} + \frac{3}{2}\omega + 4 m, \hspace{7 cm}
\end{align*}
\begin{align*}
\begin{split}
F(\textbf{R}_1, \textbf{R}_2) & = \bigg(\frac{1}{\pi d^{2}(1 + \frac{2}{3} k_f b_s d^{2})}\bigg)^{\frac{3}{2}} \text{exp}\bigg(\frac{-(1 + \frac{4}{3} k_f b_s d^{2})(\textbf{R}^{2}_{1} + \textbf{R}_{2}^{2})}{2 d^{2}}\bigg)[-\frac{E_{c}}{3} -
\frac{8 C d^{^{2}}}{3(1 + \frac{2}{3} k_f b_s d^{2})} \\& + \frac{4 m}{3} - \frac{8}{9}\overline{C} + \frac{1 + \frac{2}{3} k_f b_s d^{2}}{6 m
d^{2}}\bigg(\frac{15}{2} - (\textbf{R}^{2}_{1} + \textbf{R}_{2}^{2})(\frac{1 + \frac{2}{3} k_f b_s d^{2}}{ d^{2}})\bigg)],
\end{split}
\end{align*}
\begin{align*}
\begin{split}
F_{1}(\textbf{R}_{i},\textbf{R}_{j}) & = (\textbf{R}^{4}_{i} + \frac{15}{4(\frac{1}{2 d^{2}}+2 p)^{2}} + \frac{\textbf{R}^{2}_{i}}{\frac{1}{2
d^{2}} + 2 p})\{E_{c} - 24 \overline{C} - 28 C \textbf{R}^{2}_{j} - 28 C \frac{3}{2(\frac{1}{2 d^{2}} + 2 p)} + 2 A \\& \text{exp}(-B
\textbf{R}^{2}_{i} - B \textbf{R}^{2}_{j}) + 2 A \text{exp}(-B \textbf{R}^{2}_{i} + B \textbf{R}^{2}_{j})\} -28 C (\frac{3
\textbf{R}^{4}_{i}}{2(\frac{1}{2 d^{2}} + 2 p)} + \frac{105}{8(\frac{1}{2 d^{2}} + 2 p)^{3}} \\& + \frac{5
\textbf{R}^{2}_{i}}{2(\frac{1}{2 d^{2}} + 2 p)^{2}}) + \frac{1}{2 m}\{24 \textbf{R}^{2}_{i} - 56 p \textbf{R}^{4}_{i} + 16
p^{2}\textbf{R}^{6}_{i} - R^{4}_{i}(\frac{3}{d^{2}} + \frac{\textbf{R}^{2}_{j}}{d^{4}}) + \frac{1}{2(\frac{1}{2 d^{2}} + 2 p)}(72 - \\&
112 p \textbf{R}^{2}_{i} + 80 p^{2}\textbf{R}^{4}_{i}) + \frac{1}{\frac{1}{(2 d^{2}} + 2 p)^{2}}(400 p^{2}\textbf{R}^{2}_{i} - 840 p) +
\frac{1680 p^{2}}{8(\frac{1}{2 d^{2}} + 2 p)^{3}}\},
\end{split}
\end{align*}
\begin{align*}
E(\textbf{R}_{i}) = \text{exp}(-B \textbf{R}^{2}_{i})
\text{exp}(\frac{B^{2}\textbf{R}^{2}_{i}}{\frac{1}{2 d^{2}} + 2 p + B}), \hspace{8 cm}
\end{align*}
\begin{align*}
\begin{split}
F_{2}(\textbf{R}_{i},\textbf{R}_{j}) &= 2 \textbf{R}^{4}_{i} + \frac{10
B^{2}\textbf{R}^{2}_{j}}{(\frac{1}{2 d^{2}} + 2 p + B)^{3}} + \frac{2 B^{4}\textbf{R}^{4}_{j}}{(\frac{1}{2 d^{2}} + 2 p + B)^{4}} + \frac{15
}{2(\frac{1}{2 d^{2}} + 2 p + B)^{2}} + \frac{4 B^{2}\textbf{R}^{2}_{i}\textbf{R}^{2}_{j}}{(\frac{1}{2 d^{2}} + 2 p + B)^{2}} \\& +
\frac{2(\textbf{R}_{i}^{2})}{(\frac{1}{2 d^{2}} + 2 p + B)} - \frac{8 B^{2}(\textbf{R}_{i}.\textbf{R}_{j})^{2}}{(\frac{1}{2 d^{2}} + 2 p +
B)^{2}},
\end{split}
\end{align*}
\begin{align*}
E(\textbf{R}_{i},\textbf{R}_{j}) = (\frac{\pi}{\frac{1}{2 d^{2}} + 2 p + B})^{\frac{3}{2}} \{ 16
E(\textbf{R}_{j})F_{2}(\textbf{R}_{i},\textbf{R}_{j}) + 2 E(\textbf{R}_{i})F_{2}(\textbf{R}_{i},\textbf{R}_{i})\}, \hspace{3 cm}
\end{align*}
\begin{align*}
\begin{split}
F_{3}(\textbf{R}_{i},\textbf{R}_{j}) & = (\textbf{R}^{4}_{j} + \frac{15}{4(\frac{1}{2 d^{2}} + 2 p)^{2}} + \frac{\textbf{R}^{2}_{j}}{\frac{1}{2
d^{2}} + 2 p})\{E_{c} - 24 \overline{C} - 28 C \textbf{R}^{2}_{i} - 28 C \frac{3}{2(\frac{1}{2 d^{2}} + 2 p)} + 2 A \\& \text{exp}(-B
\textbf{R}^{2}_{i} -B \textbf{R}^{2}_{j}) + 2 A \text{exp}(-B \textbf{R}^{2}_{i} + B \textbf{R}^{2}_{j})\} - 28 C (\frac{3
\textbf{R}^{4}_{j}}{2(\frac{1}{2 d^{2}} + 2 p)} + \frac{105}{8(\frac{1}{2 d^{2}} + 2 p)^{3}} + \frac{5 \textbf{R}^{2}_{j}}{2(\frac{1}{2 d^{2}} +
2 p)^{2}}) \\& + \frac{1}{2 m} \{20 \textbf{R}^{2}_{j} - 44 p \textbf{R}^{4}_{j} + 16 p^{2}\textbf{R}^{6}_{j} - \frac{6
\textbf{R}^{4}_{j}}{d^{2}} + \frac{\textbf{R}^{4}_{j} \textbf{R}^{2}_{i}}{d^{4}} + \frac{3}{2(\frac{1}{2 d^{2}} + 2 p)}(42 - 56 p
\textbf{R}^{2}_{j} + 32 p^{2}\textbf{R}^{4}_{j} - \frac{12 R^{2}_{j}}{ d^{2}} \\& + \frac{2 R^{2}_{j}R^{2}_{i}}{d^{4}} +
\frac{R^{4}_{2}}{d^{4}}) + \frac{15}{4(\frac{1}{2 d^{2}} + 2 p)^{2}}(-12 p + 16 p^{2}\textbf{R}^{2}_{j} - \frac{6}{d^{2}} +
\frac{\textbf{R}^{2}_{j}}{d^{4}} + \frac{2 \textbf{R}^{2}_{j}}{d^{4}}) + \frac{1}{d^{4}}\frac{35\times3}{(\frac{1}{2 d^{2}} + 2 p)^{3}} + (112 p
\\& - 64 p^{2}\textbf{R}^{2}_{j} + \frac{24}{d^{2}} - \frac{4
\textbf{R}^{2}_{i}}{d^{4}})\frac{\textbf{R}^{2}_{j}}{2(\frac{1}{2 d^{2}} + 2 p)} - \frac{20 \textbf{R}^{2}_{j}}{d^{4}(\frac{1}{2 d^{2}} + 2
p)^{2}}\},
\end{split}
\end{align*}
\begin{align*}
\begin{split}
F_{4}(\textbf{R}_{i}) &= 4 m + n^{4}\{\frac{225 E_{c}}{(4 p)^{4}}(\frac{\pi}{2 p})^{3} - \frac{\pi^3}{3}\bigg(\frac{225 A}{16 (2
p+B)^{\frac{7}{2}}(2 p)^{\frac{7}{2}}} + \frac{225 \overline{c}}{16 (2 p)^{7}} + \frac{1575 c}{32 (2 p)^{8}}\bigg) + \frac{8}{2 m}\bigg(\frac{105}{(8 p)^{3}}\frac{31}{2} - \\& \frac{15}{64 p^{2}}\frac{299}{16 p} + \frac{3}{8
p}\frac{685}{256 p^{2}}- \frac{39690}{8 p^{3}} + 32 p^{2}\frac{10395}{(8 p)^{5}} + \frac{315}{p^{3}} - \frac{6615}{(8
p)^{3}}\bigg)(\frac{\pi}{4 p})^{3} + \frac{8 \nabla^{2}_{\mathbf{R}_{i}}}{2 m}\frac{225 \pi^{3}}{16384 p^{7}}\},
\end{split}
\end{align*}
\begin{align*}
\begin{split}
F_{5}(\textbf{R}_{i},\textbf{R}_{j}) &= \textbf{R}^{4}_{i}\textbf{R}^{4}_{j} + \frac{945}{(8 p)^{4}} + \frac{1}{(8 p)^{2}}(15 \textbf{R}^{4}_{i}
+ 15 \textbf{R}^{4}_{j} - 36 \textbf{R}^{2}_{i} \textbf{R}^{2}_{j} - 32(\textbf{R}_{i}.\textbf{R}_{j})^{2}) + \frac{1}{8 p}(2
\textbf{R}^{4}_{j}\textbf{R}^{2}_{i}\\& + 2 \textbf{R}^{4}_{i}\textbf{R}^{2}_{j}) + \frac{70}{(8 p)^{3}}( \textbf{R}^{2}_{i} +
\textbf{R}^{2}_{j}),
\end{split}
\end{align*}
\begin{align*}
\begin{split}
F_{6}(\textbf{R}_{i},\textbf{R}_{j}) & = (-10 A \text{exp}(-B \overline{\textbf{R}_{i} + \textbf{R}_{j}}^{2}) - 10 A \text{exp}(-B
\overline{\textbf{R}_{i} - \textbf{R}_{j}}^{2}))\{\frac{15}{(8 p)^{2}}(\textbf{R}^{4}_{j} + \textbf{R}^{4}_{i} + 4
\textbf{R}^{2}_{i}\textbf{R}^{2}_{j})\\& + \textbf{R}^{4}_{i}\textbf{R}^{4}_{j} + \frac{6}{8 p}(\textbf{R}^{6}_{i} +
\textbf{R}^{4}_{i}\textbf{R}^{2}_{j}) + \frac{210}{(8 p)^{3}}(\textbf{R}^{2}_{i} + \textbf{R}^{2}_{j}) + \frac{945}{(8 p)^{4}} -
\frac{\textbf{R}^{4}_{j}\textbf{R}^{2}_{i}}{2 p} - \frac{140 \textbf{R}^{2}_{i}}{(8 p)^{3}} - 80 \frac{\textbf{R}^{2}_{i}\textbf{R}^{2}_{i}}{(8
p)^{2}}\\& - \frac{\textbf{R}^{4}_{i}\textbf{R}^{2}_{j}}{2 p} - \frac{140 \textbf{R}^{2}_{j}}{(8 p)^{3}}
+ \frac{32(\textbf{R}_{i}.\textbf{R}_{j})^{2}}{(8 p)^{2}} + \frac{16 R^{2}_{i}R^{2}_{j}}{(8 p)^{2}}\},
\end{split}
\end{align*}
\begin{align*}
E_1(\textbf{R}_i,\textbf{R}_j)=\frac{8 \emph{f}^{a}n^{2}}{18 \sqrt{2}(2 \pi d^{2})^{\frac{3}{2}}} \text{exp}(\frac{-\textbf{R}_{j}^{2}}{2 d^{2}}) \text{exp}(-2 p \textbf{R}^{2}_{i}) [(\frac{\pi}{\frac{1}{2 d^{2}} + 2 p})^{\frac{3}{2}} F_{1}(\textbf{R}_{i},\textbf{R}_{j})- 2 A
E(\textbf{R}_{i},\textbf{R}_{j})],\hspace{1.5 cm}
\end{align*}
\begin{align*}
E_2 (\textbf{R}_i, \textbf{R}_j) = \frac{8 \emph{f}^{a}n^{2}}{18 \sqrt{2}(2 \pi d^{2})^{\frac{3}{2}}}\int
d^{3}\textbf{R}_{j} \text{exp}(\frac{-\textbf{R}_{i}^{2}}{2 d^{2}}) \text{exp}(-2 p \textbf{R}^{2}_{j}) [(\frac{\pi}{\frac{1}{2 d^{2}} + 2 p})^{\frac{3}{2}}
F_{3}(\textbf{R}_{i},\textbf{R}_{j}) - 2 A E(\textbf{R}_{2j},\textbf{R}_{i})],
\end{align*}
\begin{align*}
\begin{split}
E_3 (\textbf{R}_i, \textbf{R}_j) =& \frac{8 n^{4}\emph{f}^{a}}{3} \text{exp}(-2 p \textbf{R}_{i}^{2} - 2 p \textbf{R}_{j}^{2})\bigg[(\frac{\pi}{4 p})^{\frac{3}{2}}\big(E_{c}F_{5}(\textbf{R}_{i},\textbf{R}_{j}) + F_{6}(\textbf{R}_{i},\textbf{R}_{j})\big) + 2 A \\& \big[\text{exp}(- B \textbf{R}^{2}_{j})
\{D(\textbf{R}_{i},\textbf{R}_{j}) D_{1}(\textbf{R}_{j}) + \textbf{R}_{i}^{4}\textbf{R}_{j}^{4} + 2( \textbf{R}^{6}_{i} + \textbf{R}_{i}^{4}
\textbf{R}_{j}^{2})D_{2}(\textbf{R}_{i}) + 2(\textbf{R}^{2}_{i} + \textbf{R}^{2}_{j}) D_{3}(\textbf{R}_{j}) +
\\& D_{4}(\textbf{R}_{j}) - 4 \textbf{R}^{4}_{j} D_{5}(\textbf{R}_{i},\textbf{R}_{j}) - 4 D_{6}
(\textbf{R}_{i},\textbf{R}_{j}) - 8 \textbf{R}^{2}_{j} D_{7}(\textbf{R}_{i},\textbf{R}_{j}) - 4 \textbf{R}^{4}_{i} D_{8}(\textbf{R}_{j}) - 4
D_{9}(\textbf{R}_{j}) \\& - 8 \textbf{R}^{2}_{i} D_{10}(\textbf{R}_{j}) + 16 D_{11}(\textbf{R}_{i},\textbf{R}_{j})\}\text{exp}(\frac{(2
B)^{2}\textbf{R}^{2}_{j}}{4(B + 4 p)}) + \text{exp}(- B \textbf{R}^{2}_{i}) \{D(\textbf{R}_{i},\textbf{R}_{j}) D_{1}(\textbf{R}_{i}) +
\textbf{R}_{i}^{4}\textbf{R}_{j}^{4} \\& + 2( \textbf{R}^{6}_{i} + \textbf{R}_{i}^{4} \textbf{R}_{j}^{2})D_{2}(\textbf{R}_{i}) +
2(\textbf{R}^{2}_{i} + \textbf{R}^{2}_{j}) D_{3}(\textbf{R}_{i}) + D_{4}(\textbf{R}_{i}) - 4
\textbf{R}^{4}_{i} D_{5}(\textbf{R}_{j},\textbf{R}_{i}) - 4 D_{6}(\textbf{R}_{j},\textbf{R}_{i}) \\& - 8
\textbf{R}^{2}_{i} D_{7}(\textbf{R}_{j},\textbf{R}_{i}) - 4 \textbf{R}^{4}_{i} D_{8}(\textbf{R}_{i}) - 4 D_{9}(\textbf{R}_{i}) - 8
\textbf{R}^{2}_{j}D_{10}(\textbf{R}_{i}) + 16 D_{11}(\textbf{R}_{j},\textbf{R}_{i})\} \\& \text{exp}(\frac{(2 B)^{2}\textbf{R}^{2}_{i}}{4(B + 4 p)})\big]
+ \frac{1}{2 m}(\frac{\pi}{4 p})^{\frac{3}{2}}D_{12}(\textbf{R}_{i},\textbf{R}_{j})\bigg]
\end{split}
\end{align*}
\begin{align*}
D(\textbf{R}_{i},\textbf{R}_{j}) = \textbf{R}^{4}_{j} + \textbf{R}^{4}_{i} + 4\textbf{R}^{2}_{i}\textbf{R}^{2}_{j}, \hspace{10 cm}
\end{align*}
\begin{align*}
D_{1}(\textbf{R}_{j}) = \frac{10(2 B \textbf{R}_{j})^{2}}{(2 B+8 p)^{3}} + \frac{(2 B \textbf{R}_{j})^{4}}{(2 B + 8 p)^{4}} + \frac{15}{(2
B + 8 p)^{2}}, \hspace{7.5 cm}
\end{align*}
\begin{align*}
D_{2}(\textbf{R}_{i}) = \frac{(2 B \textbf{R}_{i})^{2}}{(2 B + 8 p)^{2}} + \frac{3}{(2B + 8 p)}, \hspace{10 cm}
\end{align*}
\begin{align*}
D_{3}(\textbf{R}_{i}) =\frac{(2 B \textbf{R}_{i})^{6}}{(2 B + 8 p)^{6}} + \frac{21 (2 B \textbf{R}_{i})^{4}}{(2 B + 8 p)^{5}} + \frac{105}{(2 B
+ 8 p)^{3}} + \frac{105(2 B \textbf{R}_{i})^{2}}{(2 B + 8 p)^{4}}, \hspace{5 cm}
\end{align*}
\begin{align*}
D_{4}(\textbf{R}_{i}) = \frac{(2 B \textbf{R}_{i})^{8}}{(2 B + 8 p)^{8}} + \frac{36 (2 B \textbf{R}_{i})^{6}}{(2 B + 8 p)^{7}} + \frac{378(2 B
\textbf{R}_{i})^{4}}{(2 B + 8 p)^{6}} + \frac{1260 (2 B \textbf{R}_{i})^{2}}{(2 B + 8 p)^{5}} + \frac{945}{(2 B + 8 p)^{4}}, \hspace{2.5 cm}
\end{align*}
\begin{align*}
D_{5}(\textbf{R}_{i},\textbf{R}_{j}) = \frac{(2 B)^{2}(\textbf{R}_{i}.\textbf{R}_{j})^{2}}{(2 B + 8 p)^{2}} + \frac{3
\textbf{R}^{2}_{i}}{(2 B + 8 p)}, \hspace{8.5 cm}
\end{align*}
\begin{align*}
\begin{split}
D_{6}(\textbf{R}_{i},\textbf{R}_{j})& = \frac{(2 B)^{6}(\textbf{R}_{i}.\textbf{R}_{j})^{2}\textbf{R}^{4}_{j}}{(2 B + 8 p)^{6}} + \frac{(2
B)^{4}}{(2 B + 8 p)^{5}}(18 \textbf{R}^{2}_{j}(\textbf{R}_{i}.\textbf{R}_{j})^{2} + \textbf{R}^{2}_{i}\textbf{R}^{4}_{j}) + \frac{(2 B)^{2}}{(2
B + 8 p)^{4}}(63(\textbf{R}_{i}.\textbf{R}_{j})^{2}\\& + 14 \textbf{R}^{2}_{i}\textbf{R}^{2}_{j} + x^{2}_{i}x^{2}_{j} + y^{2}_{i}y^{2}_{j} +
z^{2}_{i}z^{2}_{j}) + \frac{35 \textbf{R}^{2}_{i}}{(2 B + 8 p)^{3}},
\end{split}
\end{align*}
\begin{align*}
D_{7}(\textbf{R}_{i},\textbf{R}_{j}) = \frac{(2 B)^{4}}{(2 B + 8 p)^{4}}\textbf{R}_{j}^{2}(\textbf{R}_{i}.\textbf{R}_{j})^{2} + \frac{(2
B)^{2}}{(2 B + 8 p)^{3}}(7(\textbf{R}_{i}.\textbf{R}_{j})^{2} + \textbf{R}_{i}^{2}\textbf{R}_{j}^{2}) + \frac{5 \textbf{R}^{2}_{i}}{(2 B + 8 p)^{2}}, \hspace{1.5 cm}
\end{align*}
\begin{align*}
D_{8}(\textbf{R}_{j}) = \frac{(2 B)^{2}\textbf{R}_{j}^{4}}{(2 B + 8 p)^{2}} + \frac{\textbf{R}^{2}_{j}}{2 B+8 p},\hspace{9.5 cm}
\end{align*}
\begin{align*}
D_{9}(\textbf{R}_{j}) = \frac{35 \textbf{R}^{2}_{j}}{(2 B + 8 p)^{3}} + \frac{(2 B)^{6}}{(2 B + 8 p)^{6}}\textbf{R}^{8}_{j} + \frac{19 (2
B)^{4}}{(2 B + 8 p)^{5}}\textbf{R}^{6}_{j} + \frac{77(2 B)^{2}}{(2 B + 8 p)^{4}}\textbf{R}^{4}_{j},\hspace{3 cm}
\end{align*}
\begin{align*}
D_{10}(\textbf{R}_{j}) = \frac{(2 B)^{4}}{(2 B + 8 p)^{4}}\textbf{R}_{j}^{6} + \frac{(2 B)^{2}}{(2 B + 8 p)^{3}}\textbf{R}_{j}^{4} +
\frac{5}{(2 B + 8 p)^{2}}\textbf{R}_{j}^{2},\hspace{5 cm}
\end{align*}
\begin{align*}
\begin{split}
D_{11}(\textbf{R}_{i},\textbf{R}_{j})& = \frac{(2 B)^{4}}{(2 B + 8 p)^{4}}\textbf{R}_{j}^{4}(\textbf{R}_{i}.\textbf{R}_{j})^{2} + \frac{(2
B)^{2}}{(2 B + 8 p)^{3}}(5 \textbf{R}_{j}^{2}(\textbf{R}_{i}.\textbf{R}_{j})^{2} + \textbf{R}_{i}^{2}\textbf{R}_{j}^{4})\\
& + \frac{1}{(2 B + 8 p)^{2}}(2(\textbf{R}_{i}.\textbf{R}_{j})^{2}+\textbf{R}_{i}^{2}\textbf{R}_{j}^{2}),\hspace{7 cm}
\end{split}
\end{align*}
\begin{align*}
\begin{split}
D_{12}(\textbf{R}_{i},\textbf{R}_{j})& = \frac{1}{2048 p^{3}}\bigg\{- 9975 + 32768 p^{5}\textbf{R}_{i}^{4}\textbf{R}_{j}^{4}(R^{2}_{i} +
R^{2}_{j}) + 40 p(287 R_{i}^{2} + 31 R_{j}^{2})\\ & + 4096 p^{4}(2 R_{i}^{6} R_{j}^{2} - 43 R_{i}^{4} R_{j}^{4} + 2 R_{i}^{2} R_{j}^{6}) \\& + 512
p^{3} (R_{i}^{2} + R^{j}_{2})\big(15 R_{i}^{4} - 14 R_{i}^{2} R^{2}_{j} + 15 R_{j}^{4} + 16
(2(\textbf{R}_{i}.\textbf{R}_{j})^{2}+\textbf{R}_{i}^{2} \textbf{R}_{j}^{2})\big) - \\ & 64 p^{2} \bigg(287 R_{i}^{4} - 772 R_{i}^{2} R^{2}_{j} + 287
R^{4}_{j} +672\big(2(\textbf{R}_{i}.\textbf{R}_{j})^{2}+\textbf{R}_{i}^{2} \textbf{R}_{j}^{2}\big) \\& - 1024 \big(7 \textbf{R}_{i}^{2}
\textbf{R}_{j}^{2} + 14 (\textbf{R}_{i}.\textbf{R}_{j})^{2}\big)\bigg) \bigg\}.\hspace{7 cm}
\end{split}
\end{align*}
\begin{center}
\section*{APPENDIX C}
\end{center}

\subsection*{Solving the Integral Equation}
\qquad By taking the three dimensional Fourier transform of
eqs.(\ref{e12},\ref{e14}) with respect to $\textbf{R}_{1}$ and
eqs.(\ref{e13},\ref{e15}) with respect to $\textbf{R}_{2}$, these
integral equations become
\begin{align}
\begin{split}
(-E_{c}& - 8 C d^{2}-\frac{8}{3}\overline{C}+ \frac{P_{1}^{2}}{2
m}+\frac{3}{2}\omega + 4 m)\chi_{1}(\textbf{P}_{1})+ \frac{1}{(2
\pi)^{\frac{3}{2}}}\bigg(\frac{1}{\pi d^{2}(1 +\frac{2}{3} k_f b_s d^{2})}\bigg)^{\frac{3}{2}}\int
d^{3}\textbf{R}_{1}d^{3}\textbf{R}_{2} \text{exp}(\imath
\textbf{P}_{1}.\textbf{R}_{1}) \\& \text{exp}\bigg(\frac{-(1+\frac{4}{3} k_f b_s d^{2})(\textbf{R}^{2}_{1}+\textbf{R}_{2}^{2})}{2
d^{2}}\bigg)[-\frac{E_{c}}{3}-\frac{8 C d^{^{2}}}{3(1+ \frac{2}{3} k_f b_s d^{2})}+\frac{4 m}{3}-\frac{8}{9}\overline{C} +
\frac{1+ \frac{2}{3} k_f b_s d^{2}}{6 m d^{2}}\bigg(\frac{15}{2}-
(\textbf{R}^{2}_{1} + \textbf{R}_{2}^{2})\\&(\frac{1+ \frac{2}{3} k_f b_s
d^{2}}{ d^{2}})\bigg)] \chi_{2}(\textbf{R}_{2})+\frac{1}{(2
\pi)^{\frac{3}{2}}}\int d^{3}\textbf{R}_{1}d^{3}\textbf{R}_{2}
\text{exp}(\imath \textbf{P}_{1}.\textbf{R}_{1})
\text{exp}(\frac{-\textbf{\textbf{R}}_{2}^{2}}{2 d^{2}}) \text{exp}(-2 p
\textbf{R}^{2}_{1})\frac{8 \emph{f}^{a}n^{2}}{18 \sqrt{2}(2 \pi
d^{2})^{\frac{3}{2}}} \\& [(\frac{\pi}{\frac{1}{2 d^{2}}+2
p})^{\frac{3}{2}} F_{1}(\textbf{R}_{1},\textbf{R}_{2})- 2 A
E(\textbf{R}_{1},\textbf{R}_{2})]\chi^{\star}_{2}(\textbf{R}_{2})=0,
\end{split}\tag{C1}
\end{align}
\begin{align}
\begin{split}
(-E_{c}&-8 C d^{2}-\frac{8}{3}\overline{C}+ \frac{P_{2}^{2}}{2
m}+\frac{3}{2}\omega + 4 m)\chi_{2}(\textbf{P}_{2})+ \frac{1}{(2
\pi)^{\frac{3}{2}}}\bigg(\frac{1}{\pi d^{2}(1 + \frac{2}{3} k_f b_s d^{2})}\bigg)^{\frac{3}{2}}\int
d^{3}\textbf{R}_{1}d^{3}\textbf{R}_{2} \text{exp}(\imath
\textbf{P}_{2}.\textbf{R}_{2})\\& \text{exp}\bigg(\frac{-(1+ \frac{4}{3} k_f b_s d^{2})(\textbf{R}^{2}_{1}+\textbf{R}_{2}^{2})}{2
d^{2}}\bigg)[-\frac{E_{c}}{3}-\frac{8 C d^{^{2}}}{3(1+ \frac{2}{3} k_f b_s d^{2})}+\frac{4 m}{3}-\frac{8}{9}\overline{C} +
\frac{1+ \frac{2}{3} k_f b_s d^{2}}{6 m d^{2}} \bigg(\frac{15}{2}-
(\textbf{R}^{2}_{1} + \textbf{R}_{2}^{2})\\&(\frac{1+ \frac{2}{3} k_f b_s d^{2}}{ d^{2}})\bigg)]\chi_{1}(\textbf{R}_{1})+\frac{1}{(2
\pi)^{\frac{3}{2}}}\int d^{3}\textbf{R}_{1}d^{3}\textbf{R}_{2}
\text{exp}(\imath \textbf{P}_{2}.\textbf{R}_{2})
\text{exp}(\frac{-\textbf{\textbf{R}}_{1}^{2}}{2 d^{2}}) \text{exp}(-2 p
\textbf{R}^{2}_{2})\frac{8 \emph{f}^{a}n^{2}}{18 \sqrt{2}(2 \pi
d^{2})^{\frac{3}{2}}}\\& [(\frac{\pi}{\frac{1}{2 d^{2}}+2
p})^{\frac{3}{2}}
F_{1}(\textbf{R}_{2},\textbf{R}_{1})- 2 A E(\textbf{R}_{2},\textbf{R}_{1})]\chi^{\star}_{1}(\textbf{R}_{1})=0,
\end{split}\tag{C2}
\end{align}
\begin{align}
\begin{split}
\bigg(4 m & + n^{4}\big\{\frac{225 E_{c}}{(4 p)^{4}}(\frac{\pi}{2
p})^{3}-\frac{\pi^3}{3}\bigg(\frac{225 A}{16 (2
p+B)^{\frac{7}{2}}(2 p)^{\frac{7}{2}}} + \frac{225 \overline{c}}{16 (2 p)^{7}} + \frac{1575 c}{32 (2 p)^{8}}\bigg)+\frac{8}{2 m}(\frac{105}{(8
p)^{3}}\frac{31}{2}-\frac{15}{64 p^{2}}\frac{299}{16 p}\\& + \frac{3}{8
p}\frac{685}{256 p^{2}}- \frac{39690}{8 p^{3}}+32
p^{2}\frac{10395}{(8 p)^{5}} +
\frac{315}{p^{3}}-\frac{6615}{p^{4}})(\frac{\pi}{4 p})^{3}-\frac{8
P^{2}_{1s} }{2 m}\frac{225 \pi^{3}}{16384
p^{7}}\big\}\bigg)\chi^{\star}_{1}(\textbf{P}_{1s})\\& + \frac{1}{(2
\pi)^{\frac{3}{2}}}\frac{8 \emph{f}^{a}n^{2}}{18 \sqrt{2}(2 \pi
d^{2})^{\frac{3}{2}}}\int d^{3}\textbf{R}_{1} d^{3}\textbf{R}_{2}
\text{exp}(\imath
\textbf{P}_{1s}.\textbf{R}_{1})\text{exp}(\frac{-\textbf{R}_{1}^{2}}{2
d^{2}}) \text{exp} (- 2 p
\textbf{R}_{2}^{2})[......]\chi_{2}(\textbf{R}_{2})\\ & + \frac{1}{(2
\pi)^{\frac{3}{2}}} \frac{8 \emph{f}^{a} n^{4}}{3}\int
d^{3}\textbf{R}_{1} d^{3}\textbf{R}_{2} \text{exp}(\imath
\textbf{P}_{11}.\textbf{R}_{1}) \text{exp}(-2 p\textbf{R}_{1}^{2})\text{exp}(-2
p\textbf{R}_{2}^{2})[......]\chi^{\star}_{2}(\textbf{R}_{2})= 0,
\end{split}\tag{C3}
\end{align}
\begin{align}
\begin{split}
\bigg(4 m & + n^{4}\big\{\frac{225 E_{c}}{(4 p)^{4}}(\frac{\pi}{2
p})^{3}-\frac{\pi^3}{3}\bigg(\frac{225 A}{16 (2
p+B)^{\frac{7}{2}}(2 p)^{\frac{7}{2}}} + \frac{225 \overline{c}}{16 (2 p)^{7}} + \frac{1575 c}{32 (2 p)^{8}}\bigg) + \frac{8}{2 m}(\frac{105}{(8
p)^{3}}\frac{31}{2}-\frac{15}{64 p^{2}}\frac{299}{16 p}\\ & +\frac{3}{8
p}\frac{685}{256 p^{2}} - \frac{39690}{8 p^{3}}+32
p^{2}\frac{10395}{(8 p)^{5}}+
\frac{315}{p^{3}}-\frac{6615}{p^{4}})(\frac{\pi}{4 p})^{3}-\frac{8
P^{2}_{2s} }{2 m}\frac{225 \pi^{3}}{16384
p^{7}}\big\}\bigg)\chi^{\star}_{2}(\textbf{P}_{2s})\\& + \frac{1}{(2
\pi)^{\frac{3}{2}}}\frac{8 \emph{f}^{a}n^{2}}{18 \sqrt{2}(2 \pi
d^{2})^{\frac{3}{2}}}\int d^{3}\textbf{R}_{1} d^{3}\textbf{R}_{2}
\text{exp}(\imath
\textbf{P}_{2s}.\textbf{R}_{2})\text{exp}(\frac{-\textbf{R}_{2}^{2}}{2
d^{2}}) \text{exp} (- 2 p
\textbf{R}_{1}^{2})[......]\chi_{1}(\textbf{R}_{1}) \\& +\frac{1}{(2
\pi)^{\frac{3}{2}}} \frac{8 \emph{f}^{a} n^{4}}{3}\int
d^{3}\textbf{R}_{1} d^{3}\textbf{R}_{2} \text{exp}(\imath
\textbf{P}_{22}.\textbf{R}_{2}) \text{exp}(-2 p\textbf{R}_{1}^{2} - 2
p\textbf{R}_{2}^{2})[......]\chi^{\star}_{1}(\textbf{R}_{1})= 0,
\end{split}\tag{C4}
\end{align}
where $\chi(\textbf{P})$ is the fourier transform of
$\chi(\textbf{R})$. $P_{1}$,$P_{2}$ are conjugate to $R_{1}$,
$R_{2}$ for ground state gluonic field and $P_{1s}$, $P_{2s}$ are
conjugate to $R_{1}$, $R_{2}$ for excited state gluonic field. The
off-diagonal terms of eq.(C3) and (C4) are too lengthy and not used
for results, so dots are used to reduce the length of equations. The
above eqs.(C1-C4) have formal solutions as ~\cite{B. Masud91}
\begin{align}
\begin{split}
\chi_{1}(\textbf{P}_{1})&=\delta(P_{1}-P_{c}(1))/P^{2}_{c}(1)-\frac{1}{\triangle_{1}(P_{1})}
\frac{1}{(2 \pi)^{\frac{3}{2}}}\bigg(\frac{1}{\pi d^{2}(1 + \frac{2}{3} k_f b_s d^{2})}\bigg)^{\frac{3}{2}}\int
d^{3}\textbf{R}_{1}d^{3}\textbf{R}_{2} \quad \text{exp}(\imath
\textbf{P}_{1}.\textbf{R}_{1}) \\& \text{exp}\bigg(\frac{-(1+ \frac{4}{3} k_f b_s d^{2})(\textbf{R}^{2}_{1}+\textbf{R}_{2}^{2})}{2
d^{2}}\bigg)[-\frac{E_{c}}{3}-\frac{8 C d^{^{2}}}{3(1+ \frac{2}{3} k_f b_s d^{2})}+\frac{4 m}{3}-\frac{8}{9}\overline{C} +
\frac{1+ \frac{2}{3} k_f b_s d^{2}}{6 m d^{2}}\bigg(\frac{15}{2}-
(\textbf{R}^{2}_{1} + \textbf{R}_{2}^{2})\\&(\frac{1+ \frac{2}{3} k_f b_s
d^{2}}{
d^{2}})\bigg)]\chi_{2}(\textbf{R}_{2})-\frac{1}{\triangle_{1}(P_{1})}\frac{1}{(2
\pi)^{\frac{3}{2}}}\int d^{3}\textbf{R}_{1}d^{3}\textbf{R}_{2}
\text{exp}(\imath \textbf{P}_{1}.\textbf{R}_{1}) \quad \text{exp}(\frac{-
R_{2}^{2}}{2 d^{2}})\quad \text{exp}(-2 p R^{2}_{1})\\& \frac{8
\emph{f}^{a}n^{2}}{18 \sqrt{2}(2 \pi d^{2})^{\frac{3}{2}}}
[(\frac{\pi}{\frac{1}{2 d^{2}}+2 p})^{\frac{3}{2}}
F_{1}(\textbf{R}_{1},\textbf{R}_{2})- 2 A
E(\textbf{R}_{1},\textbf{R}_{2})]\chi^{\star}_{2}(\textbf{R}_{2}),
\end{split}\tag{C5}
\end{align}
\begin{align}
\begin{split}
\chi_{2}(\textbf{P}_{2})&=\delta(P_{2}-P_{c}(2))/P^{2}_{c}(2)-\frac{1}{\triangle_{2}(P_{2})}
\frac{1}{(2 \pi)^{\frac{3}{2}}}\bigg(\frac{1}{\pi d^{2}(1 + \frac{2}{3} k_f b_s d^{2})}\bigg)^{\frac{3}{2}}\int
d^{3}\textbf{R}_{1}d^{3}\textbf{R}_{2}\quad \text{exp}(\imath
\textbf{P}_{2}.\textbf{R}_{2})\\& \text{exp}\bigg(\frac{-(1+ \frac{4}{3} k_f b_s d^{2})(\textbf{R}^{2}_{1} + \textbf{R}_{2}^{2})}{2
d^{2}}\bigg)[-\frac{E_{c}}{3}-\frac{8 C d^{^{2}}}{3(1+ \frac{2}{3} k_f b_s d^{2})}+\frac{4 m}{3}-\frac{8}{9}\overline{C}+\frac{1+ \frac{2}{3} k_f b_s d^{2}}{6 m d^{2}}\bigg(\frac{15}{2}-
(\textbf{R}^{2}_{1} + \textbf{R}_{2}^{2})\\&(\frac{1+ \frac{2}{3} k_f b_s
d^{2}}{
d^{2}})\bigg)]\chi_{1}(\textbf{R}_{1})-\frac{1}{\triangle_{2}(P_{2})}\frac{1}{(2
\pi)^{\frac{3}{2}}}\int d^{3}\textbf{R}_{1}d^{3}\textbf{R}_{2}
\text{exp}(\imath \textbf{P}_{2}.\textbf{R}_{2})\quad
\text{exp}(\frac{-R_{1}^{2}}{2 d^{2}})\quad \text{exp}(-2 p R^{2}_{2})\\& \frac{8
\emph{f}^{a}n^{2}}{18 \sqrt{2}(2 \pi d^{2})^{\frac{3}{2}}}
[(\frac{\pi}{\frac{1}{2 d^{2}}+2 p})^{\frac{3}{2}}
F_{1}(\textbf{R}_{2},\textbf{R}_{1})- 2 A
E(\textbf{R}_{2},\textbf{R}_{1})]\chi^{\star}_{1}(\textbf{R}_{1}),
\end{split}\tag{C6}
\end{align}
\begin{equation}
\begin{split}
\chi^{\star}_{1}(\textbf{P}_{1s})&=\delta(P_{1s}-P_{c}(1s))/P^{2}_{c}(1s)-\frac{1}{\triangle_{1}(P_{1s})}
\frac{1}{(2 \pi)^{\frac{3}{2}}}\frac{8 \emph{f}^{a}n^{2}}{18
\sqrt{2}(2 \pi d^{2})^{\frac{3}{2}}}\int d^{3}\textbf{R}_{1}
d^{3}\textbf{R}_{2} \text{exp}(\imath \textbf{P}_{1s}.\textbf{R}_{1})\\&
\text{exp}(\frac{-R_{1}^{2}}{2 d^{2}}) \text{exp} (- 2 p
R_{2}^{2})[......]\chi_{2}(\textbf{R}_{2})-\frac{1}{\triangle_{1}(P_{1s})}\frac{8
\emph{f}^{a} n^{4}}{3}\frac{1}{(2 \pi)^{\frac{3}{2}}}\int
d^{3}\textbf{R}_{1} d^{3}\textbf{R}_{2} \text{exp}(\imath
\textbf{P}_{1s}.\textbf{R}_{1})\\& \text{exp}(-2 p R_{1}^{2})\text{exp}(-2 p
R_{2}^{2})[......]\chi^{\star}_{2}(\textbf{R}_{2}),
\end{split}\tag{C7}
\end{equation}
\begin{align}
\begin{split}
\chi^{\star}_{2}(\textbf{P}_{2s})& =\delta(P_{2s}-P_{c}(2s))/P^{2}_{c}(2s)-\frac{1}{\triangle_{2}(P_{2s})} \frac{1}{(2 \pi)^{\frac{3}{2}}}\frac{8
\emph{f}^{a}n^{2}}{18 \sqrt{2}(2 \pi d^{2})^{\frac{3}{2}}}\int d^{3}\textbf{R}_{1} d^{3}\textbf{R}_{2} \text{exp}(\imath
\textbf{P}_{2s}.\textbf{R}_{2})\\& \text{exp}(\frac{-R_{2}^{2}}{2 d^{2}}) \text{exp} (- 2 p
R_{1}^{2})[......]\chi_{1}(\textbf{R}_{1})-\frac{1}{\triangle_{2}(Pf_{2s})}\frac{8 \emph{f}^{a} n^{4}}{3}\frac{1}{(2 \pi)^{\frac{3}{2}}}\int
d^{3}\textbf{R}_{1} d^{3}\textbf{R}_{2} \text{exp}(\imath \textbf{P}_{2s}.\textbf{R}_{2})\\& \text{exp}(-2 p R_{2}^{2})\text{exp}(-2 p
R_{1}^{2})[......]\chi^{\star}_{1}(\textbf{R}_{1}),
\end{split}\tag{C8}
\end{align}
where
\begin{equation}
\begin{split}
& \triangle_{1}(\textbf{P}_{i})= (-E_{c}- 8 C
d^{2}-\frac{8}{3}\overline{C}+ \frac{\textbf{P}_{i}^{2}}{2
m}+\frac{3}{2}\omega + 4 m)-\iota\varepsilon,\\&
\triangle_{2}(\textbf{P}_{is})=4 m + n^{4}\{\frac{225 E_{c}}{(4
p)^{4}}(\frac{\pi}{2 p})^{3}-\frac{\pi^3}{3}\bigg(\frac{225 A}{16 (2
p+B)^{\frac{7}{2}}(2 p)^{\frac{7}{2}}} + \frac{225 \overline{c}}{16 (2 p)^{7}} + \frac{1575 c}{32 (2 p)^{8}}\bigg) +
\frac{4}{m}\bigg(\frac{105}{(8 p)^{3}}\frac{31}{2}- \\& \qquad \frac{15}{64
p^{2}}\frac{299}{16 p}+\frac{3}{8 p}\frac{685}{256 p^{2}} -
\frac{39690}{8 p^{3}}+32 p^{2}\frac{10395}{(8 p)^{5}}+
\frac{315}{p^{3}}-\frac{6615}{(8 p)^{3}}\bigg)(\frac{\pi}{4
p})^{3}-\frac{ 8 (\textbf{P}_{is} )^{2}}{2 m}\frac{225
\pi^{3}}{16384 p^{7}}\}-\iota\varepsilon,
\end{split}\tag{C9}
\end{equation}
and
\begin{equation}
\begin{split}
&P_{c}(2) = P_{c}(1)=\sqrt{2 m(E_{c}- 4 m - \frac{3}{m d^{2}} + \frac{8}{3} \overline{C})},\\&
P_{c}(2s) = P_{c}(1s)=
\sqrt{2 m(E_{c} - 4 m)- \big(0.6667 \overline{C}+\frac{1.1667 C}{p}+\frac{13.682 p^{7/2}}{(0.0657 + 2 p)^{7/2}} \big) m- 64 p}.
\end{split}\tag{C10}
\end{equation}
Using the Born approximation, the integration on $\textbf{R}_{1}$ and
$\textbf{R}_{2}$ in eq.(C5) can be performed to give
\begin{equation}
\begin{split}
\chi_{1}(\textbf{P}_{1})& =\delta(P_{1}-P_{c}(1))/P^{2}_{c}(1)-
\frac{1}{\triangle_{1}(P_{1})}\bigg[\frac{1}{3}(\frac{2}{\pi})^{\frac{1}{2}}(\frac{1}{2 \pi})^{\frac{3}{2}}\bigg(\frac{1}{\pi d^{2}(1 + \frac{2}{3} k_f b_s d^{2})}\bigg)^{\frac{3}{2}} \bigg(\frac{2 \pi d^{2}}{1+ \frac{4}{3} k_f b_s d^{2}}\bigg)^{3} \\ & \text{exp}\bigg(-\frac{(p^{2}_{1}+
p^{2}_{2})d^{2}}{2(1+ \frac{4}{3} k_f b_s d^{2})}\bigg) \Omega_{1}\bigg]
-\frac{1}{\triangle_{1}(P_{1})}\frac{1}{3}(\frac{2}{\pi})^\frac{1}{2}(\frac{1}{\pi d^2})^\frac{3}{2}f^{a }n^2(\frac{1}{2
\pi})^\frac{3}{2}\bigg[\text{exp}(\frac{-P^{2}_{1s}d^{2}}{2})(2 \pi d^{2})^{\frac{3}{2}} \\& \bigg((\frac{\pi}{2 p+\frac{1}{2
d^{2}}})^{\frac{3}{2}}F(p)-\frac{4 A}{6\sqrt{2}}F_{1}(p)\bigg) - \frac{4 A}{6 \sqrt{2}}(\frac{\pi}{2p+ \frac{1}{2
d^{2}}})^{\frac{3}{2}}\text{exp}(\frac{-p^{2}_{1}}{8 p+4 B})(\frac{\pi}{2 p +B})^{\frac{3}{2}}(\frac{\pi}{F_{2}(p)})^{\frac{3}{2}} \\& \text{exp}\big(\frac{(\iota
P_{1s}-4 \iota \frac{P_{1}B}{8p +4 B})^{2}}{4 F_{2}(p)}\big)\Omega_{2} - \frac{4 A}{6 \sqrt{2}}(\frac{\pi}{2p+ \frac{1}{2
d^{2}}})^{\frac{3}{2}} \text{exp}(\frac{-P^{2}_{1}}{8 p+4 B})(\frac{\pi}{2 p +B})^{\frac{3}{2}}(\frac{\pi}{F_{2}(p)})^{\frac{3}{2}}\text{exp}(\frac{(\iota
P_{1s}+4\iota \frac{P_{1}B}{8p +4 B})^{2}}{4 F_{2}(p)})\Omega_{3} \\& + \frac{16 A}{3 \sqrt{2}}(\frac{\pi}{F_{3}(p)})^{\frac{3}{2}}
\text{exp}(\frac{-P^{2}_{1}}{8 p})(\frac{\pi}{2 p})^{\frac{3}{2}}\Omega_{4}\bigg] (\frac{\pi}{B+\frac{1}{2
d^{2}}-\frac{B^{2}}{F_{3}(p)}})^{\frac{3}{2}} \text{exp}(\frac{-P^{2}_{1s}}{4(B+\frac{1}{2 d^{2}}-\frac{B^{2}}{F_{3}(p)})}).
\end{split}\tag{C11}
\end{equation}
Here,
\begin{equation*}
\Omega_{1}= -E_{c}-\frac{8 C d^{^{2}}}{(1+ \frac{2}{3} k_f b_s d^{2})}+ \frac{8}{3} C + 4 m + \frac{15(1+ \frac{2}{3} k_f b_s
d^{2})}{4 m d^{2}}-(\frac{(1 + \frac{2}{3} k_f b_s d^{2})^{2}}{2 m
d^{4}})(\frac{-P^{2}_{1} d^{4}-P^{2}_{2} d^{4}}{(1 + \frac{4}{3} k_f b_s
d^{2})^{2}} + \frac{6 d^{2}}{1+ \frac{4}{3} k_f b_s d^{2}})
\end{equation*}
\begin{equation*}
\begin{split}
\Omega_{2} = &\bigg(\frac{-5 P^{2}_{1}}{4 (2 p+B)^{3}} + \frac{15}{4(2 p+B)^{2}}+\frac{15}{4(\frac{1}{2 d^{2}}+2 p)^{2}} - \frac{P^{2}_{1}}{4(2
p+B)^{2}(\frac{1}{2 d^{2}}+2 p)} + \frac{3}{2(2 p+B)(\frac{1}{2 d^{2}} + 2p)}\bigg) + \\& \bigg(\frac{5B^{2}}{(2p+B)^{3}}+
\frac{4B^{2}}{4(2p+B)^{2}(\frac{1}{2 d^{2}} + 2 p)}\bigg)\bigg(\frac{\iota P_{1s} - \frac{4\iota P_{1}B}{8p + 4 B}}{4 F^{2}_{2}(p)} + \frac{3}{2
F_{2}(p)}\bigg) \\& + \bigg(\frac{-10\iota B}{4 (2p + B)^{3}} + \frac{2\iota B}{4 (2p + B)^{2}(\frac{1}{2 d^{2}} + 2p)}\bigg)\bigg(\frac{-2\iota
P^{2}_{1}B}{4 (2p + B)} + \iota P_{1}.P_{1s}\bigg)\frac{1}{2 F_{2}(p)} \\& + \frac{1}{16(2p + B)^{4}} \bigg\{P^{4}_{1} + 16 B^{4}\bigg(\frac{5(\iota P_{1s}-
\frac{4\iota P_{1}B}{8p + 4 B})^{2}}{4 F^{3}_{2}(p)} + \frac{(\iota P_{1s}- \frac{4\iota P_{1}B}{8p + 4 B})^{4}}{16 F^{4}_{2}(p)} +
\frac{15}{4F^{2}_{2}(p)}\bigg) \\& - 8 B^{2}P^{2}_{1}\bigg(\frac{(\iota P_{1s}- \frac{4\iota P_{1}B}{8p + 4
B})^{2}}{4F^{2}_{2}(p)} + \frac{3}{2F_{2}(p)}\bigg) + 8 B P^{2}_{1}\iota \bigg(\frac{-2 \iota P^{2}_{1}\frac{B}{8p + 4 B} + \iota
P_{1}.P_{1s}}{2F_{2}(p)}\bigg) \\& - 32\iota B^{3}
(P_{1x}+P_{1y}+P_{1z})\Bigg(\bigg(\frac{\iota B^{3}P^{3}_{1}}{(2
p+B)^{3}}-\frac{B^{2}\iota}{(2p+B)^{2}}P^{2}_{1}(P_{1sx}+P_{1sy}+P_{1sz}) - 64\frac{B^{2}\iota}{(8 p + 4 B)^{2}}P_{1}(P_{1}.P_{1s}) \\& +\frac{16
\iota B}{8p+ 4B} (P_{1}.P_{1s})P_{1s} + \frac{4 \iota B}{8p + 4B}(P_{1x} + P_{1y} + P_{1z})P^{2}_{1s} -\iota P^{3}_{1s}\bigg)\frac{1}{8
F_{2}(p)^{3}} + \frac{(\iota P_{1s} - 4 \iota \frac{P_{1}B}{8p + 4 B})}{F_{2}(p)^{2}}\Bigg) \\& -16 B^{2}\bigg(P^{2}_{1}\frac{(\iota P_{1s}-4 \iota
\frac{P_{1}B}{8p +4 B})^{2}}{4 F_{2}(p)^{2}}+\frac{P^{2}_{1}}{2 F_{2}(p)}\bigg)\bigg\}
\end{split}
\end{equation*}
\begin{equation*}
\begin{split}
\Omega_{3} =& \quad\bigg(\frac{-5 P^{2}_{1}}{4 (2 p+B)^{3}}\quad + \quad\frac{15}{4(2 p+B)^{2}}\quad + \quad\frac{15}{4(\frac{1}{2 d^{2}}\quad +
\quad2 p)^{2}}-\frac{P^{2}_{1}}{4(2 p+B)^{2}(\frac{1}{2 d^{2}}+2 p})\\&+\frac{3}{2(2 p+B)(\frac{1}{2
d^{2}}+2p)}\bigg)+\bigg(\frac{5B^{2}}{(2p+B)^{3}}+\frac{4B^{2}}{4(2p+B)^{2}(\frac{1}{2 d^{2}}+2 p)}\bigg)\bigg(\frac{\iota P_{1s}+ \frac{4\iota
P_{1}B}{8p+4 B}}{4 F^{2}_{2}(p)}+\frac{3}{2 F_{2}(p)}\bigg)\\&+\bigg(\frac{-10\iota B}{4 (2p+B)^{3}}+\frac{2\iota B}{4 (2p+B)^{2}(\frac{1}{2
d^{2}}+2p)}\bigg)\bigg(\frac{-2\iota P^{2}_{1}B}{4 (2p+B)}+\iota P_{1}.P_{1s}\bigg)\frac{1}{2
F_{2}(p)}+\frac{1}{16(2p+B)^{4}}\bigg\{P^{4}_{1}\\& +16 B^{4}\bigg(\frac{5(\iota P_{1s}+ \frac{4\iota P_{1}B}{8p+4 B})^{2}}{4
F^{3}_{2}(p)}+\frac{(\iota P_{1s}+ \frac{4\iota P_{1}B}{8p+4 B})^{4}}{16 F^{4}_{2}(p)}+\frac{15}{4F^{2}_{2}(p)}\bigg)-8
B^{2}P^{2}_{1}\bigg(\frac{(\iota P_{1s}+ \frac{4\iota P_{1}B}{8p+4 B})^{2}}{4F^{2}_{2}(p)}+\frac{3}{2F_{2}(p)}\bigg)\\& -8 B
P^{2}_{1}\iota\bigg(\frac{2 \iota P^{2}_{1}\frac{B}{8p+4 B}+\iota P_{1}.P_{1s}}{2F_{2}(p)}\bigg) + 32\iota
B^{3}(P_{1x}+P_{1y}+P_{1z})\Bigg(\bigg(\frac{-\iota B^{3}P^{3}_{1}}{(2 p+B)^{3}}-\frac{B^{2}\iota}{(2p+B)^{2}}P^{2}_{1}(P_{1sx}\\& + P_{1sy} +
P_{1sz})-64\frac{B^{2}\iota}{(8 p+4 B)^{2}}P_{1}(P_{1}.P_{1s})-\frac{16 \iota B}{8p+ 4B} (P_{1}.P_{1s})P_{1s}-\frac{4 \iota B}{8p
+4B}(P_{1x}+P_{1y}+P_{1z})P^{2}_{1s}\\&-\iota P^{3}_{1s}\bigg)\frac{1}{8 F_{2}(p)^{3}}+\frac{(\iota P_{1s}+4 \iota \frac{P_{1}B}{8p +4
B})}{F_{2}(p)^{2}}\Bigg)-16 B^{2}\bigg(P^{2}_{1}\frac{(\iota P_{1s}+4 \iota \frac{P_{1}B}{8p +4 B})^{2}}{4 F_{2}(p)^{2}}+\frac{P^{2}_{1}}{2
F_{2}(p)}\bigg)\bigg\}
\end{split}
\end{equation*}
\begin{equation*}
\begin{split}
\Omega_{4}=&\frac{15}{2F^{2}_{3}(p) }-\frac{10 P^{2}_{1}}{32 p^{3}}+\frac{2 P^{4}_{1}}{(4 p)^{4}}+\frac{30}{16
p^{2}}-\frac{2}{F_{3}(p)}(\frac{-P^{2}_{1}}{16 p^{2}}+\frac{3}{4 p})+\bigg(\frac{10 B^{2}}{F^{3}_{3}(p)}-\frac{8 B^{2}}{4 p
F^{2}_{3}(p)}\\&(1-\frac{P^{2}_{1}}{4 p})+\frac{4 B^{2}}{F^{2}_{3}(p)}(\frac{-P^{2}_{1}}{16 p^{2}}+\frac{3}{4 p})\bigg)\bigg(\frac{P^{2}_{1s}}{4
B-4 \frac{B^{2}}{F_{3}(p)}}+\frac{3}{2 B-2 \frac{B^{2}}{F_{3}(p)}}\bigg)+ \frac{2 B^{4}}{F^{4}_{3}(p)}\bigg(\frac{5 P^{2}_{1s}}{4(B-
\frac{B^{2}}{F_{3}(p)})^{3}}\\&+\frac{P^{4}_{1s}}{16 (B-\frac{B^{2}}{F_{3}(p)})^{4}} +\frac{15}{4(B-\frac{B^{2}}{F_{3}(p)})^{2}})
\end{split}
\end{equation*}
\begin{equation*}
\begin{split}
&F(p)=\bigg\{\bigg(\frac{-5 P^{2}_{1}}{32 p^{3}} +
\frac{P^{4}_{1}}{16 (2 p)^{4}} + \frac{15}{16 p^{2}}\bigg) \bigg(E +
\frac{24 \overline{c}}{6 \sqrt{2}} + \frac{42 c}{6
\sqrt{2}(\frac{1}{2 d^{2}} + 2 p)} - \frac{28 p}{m} - \frac{3}{2 m
d^{2}} + \frac{20 p^{2}}{m (\frac{1}{2 d^{2}} + 2 p)}\\& \qquad -
\frac{8 c}{6\sqrt{2}(\frac{1}{2 d^{2}} + 2 p )} + (\frac{28 c}{6
\sqrt{2}} + \frac{1}{2 m d^{4}})(-P_{11}^{2} d^{4} + 3 d^{2})\bigg)
+ \bigg(\frac{15 E}{4(\frac{1}{2 d^{2}} + 2 p)^{2}} + \frac{420
c}{24\sqrt{2}(\frac{1}{2 d^{2}} + 2p)^{2}}\\& \qquad (-P_{1s}^{2}
d^{4} + 3 d^{2}) + \frac{90 \overline{c}}{6 \sqrt{2}(\frac{1}{2
d^{2}} + 2 p)^{2}} + \frac{2940 c}{48 \sqrt{2}(\frac{1}{2 d^{2}} + 2
p)^{3}} + \frac{72}{4 m (\frac{1}{2 d^{2}} + 2 p)} - \frac{105 p}{m
(\frac{1}{2 d^{2}} + 2 p)^{2}}\\& \qquad + \frac{105 p^{2}}{m
(\frac{1}{2 d^{2}} + 2 p)^{3}}\bigg) + \bigg(\frac{-P^{2}_{1}}{16
p^{2}} + \frac{3}{4 p}\bigg)\bigg(\frac{E}{(\frac{1}{2 d^{2}} + 2
p)} - \frac{28 c}{6\sqrt{2}(\frac{1}{2 d^{2}} + 2 p )}(-P_{11}^{2}
d^{4} + 3 d^{2}) + \frac{12}{m}\\& \qquad + \frac{24 \overline{c}}{6
\sqrt{2}(\frac{1}{2 d^{2}} + 2 p)} -\frac{120
c}{24\sqrt{2}(\frac{1}{2 d^{2}} + 2 p)^{2}} + \frac{70 c}{6
\sqrt{2}(\frac{1}{2 d^{2}} + 2 p)^{2}} - \frac{28
p}{m(\frac{1}{2d^{2}} + 2p)} + \frac{50 p^{2}}{m(\frac{1}{2d^{2}} +
2p)^{2}}\bigg)\\& \qquad + \bigg(\frac{8 p^{2}}{m} - \frac{8
c}{6\sqrt{2}}\bigg)\bigg(-\frac{P_{1}^{6}}{(4 p)^{6}} + \frac{21
P^{4}_{1}}{(4 p)^{5}} + \frac{105}{(4 p)^{3}} - \frac{105
P_{1}^{2}}{(4 p)^{4}}\bigg)\bigg\} (\frac{\pi}{2
p})^{\frac{3}{2}}\text{exp}(\frac{-P^{2}_{1}}{8 p}),\\&
F_{1}(p)=(\frac{\pi}{\frac{1}{2 d^{2}}+2
p+B})^{\frac{3}{2}}(\frac{\pi}{2 p+B-\frac{B^{2}}{(\frac{1}{2
d^{2}}+2p+B)}})^{\frac{3}{2}}\text{exp}(\frac{-P^{2}_{1}}{4(2
p+B-\frac{B^{2}}{(\frac{1}{2 d^{2}}+2p+B)})})\\& \qquad \{(2+\frac{2
B^{4}}{(\frac{1}{2 d^{2}}+2 p+B)^{4}}-\frac{4 B^{2}}{(\frac{1}{2
d^{2}}+2 p+B)^{2}})(\frac{-5 P^{2}_{1}}{4(2
p+B-\frac{B^{2}}{(\frac{1}{2
d^{2}}+2p+B)})^{3}}+\frac{P^{4}_{1}}{16(2
p+B-\frac{B^{2}}{(\frac{1}{2 d^{2}}+2p+B)})^{4}}\\& \qquad +
\frac{15}{4(2 p+B-\frac{B^{2}}{(\frac{1}{2
d^{2}}+2p+B)})^{2}})+(\frac{10 B^{2}}{(\frac{1}{2 d^{2}}+2
p+B)^{3}}+\frac{2}{\frac{1}{2 d^{2}}+2 p+B}) (\frac{- P^{2}_{1}}{4(2
p+B-\frac{B^{2}}{(\frac{1}{2 d^{2}}+2p+B)})^{2}}\\& \qquad +
\frac{3}{2(2 p+B-\frac{B^{2}}{(\frac{1}{2
d^{2}}+2p+B)})})+\frac{15}{2(\frac{1}{2 d^{2}}+2 p+B)^{2}}\}, \quad \textrm{with}
\end{split}
\end{equation*}
$F_{2}(p)=\frac{1}{2 d^{2}}+B-\frac{4 B^{2}}{4(2 p+B)}$,\quad and
\qquad
$F_{3}(p)=\frac{1}{2 d^{2}}+2 p+B.$\\
Now from this eq.(C11)
the elements of transition matrix (T) can be
found. As shown in ref.~\cite{P. PennanenB}, the gluonic excitations
are orthogonal to the ground states i.e.
\begin{align*}
\langle 1 \mid1^{\star}\rangle= \langle 1^{\star}\mid 1\rangle = \langle 2\mid 2^{\star}\rangle = \langle 2^{\star}\mid 2\rangle = 0.
\end{align*}
This gives
\begin{align*}
T_{11^{\star}}=T_{1^{\star}1}=T_{22^{\star}}=T_{2^{\star}2}=0.
\end{align*}
 Thus the only possible transition amplitudes are $T_{11}$, $T_{12}$, $T_{21}$,
$T_{22}$, $T_{12^{\star}}$ and $T_{21^{\star}}$. As the eq.(C5) and
(C6) are similar (interchanging $\textbf{R}_{1}$ and
$\textbf{R}_{2}$ in eq.(C5) gives eq.(C6)), so
\begin{align*}
\begin{split}
T_{11}=T_{22},\\
T_{12}=T_{21},
\end{split}
\end{align*}
\textrm{and}
\begin{align*}
T_{12^{\star}}=T_{21^{\star}}.
\end{align*}

\end{document}